\documentclass[10pt,twocolumn,article]{IEEEtran}
\usepackage{times}
\usepackage{amsbsy}
\usepackage{latexsym}
\usepackage{amssymb,bm}
\usepackage{mathtools}
\usepackage{amsmath,amssymb,amsfonts,epsfig,graphicx}
\usepackage[ansinew]{inputenc}
\DeclareMathOperator{\Tr}{Tr}

\title{\huge Multi-Step Knowledge-Aided Iterative ESPRIT \\ for Direction Finding  \\
\author{Silvio F. B. Pinto $^1$ and Rodrigo C. de Lamare $^{1,2}$ \\
Center for Telecommunications Studies (CETUC) \\ $^1$ Pontifical
Catholic
University of Rio de Janeiro, RJ, Brazil.\\
$^2$ Department of Electronics, University of York, UK \\
Emails: silviof@cetuc.puc-rio.br, delamare@cetuc.puc-rio.br}}
\begin{document}
\maketitle
\begin{abstract}
In this work, we propose a subspace-based algorithm for DOA
estimation which iteratively reduces the disturbance factors of the
estimated data covariance matrix and incorporates prior knowledge
which is gradually obtained on line. An analysis of the MSE of the
reshaped data covariance matrix is carried out along with
comparisons between computational complexities of the proposed and
existing algorithms. Simulations focusing on closely-spaced sources,
where they are uncorrelated and correlated, illustrate the
improvements achieved.
\end{abstract}


\section{Introduction}
\label{introduction}

In array signal processing, direction-of-arrival (DOA) estimation is
a key task in a broad range of important applications including
radar and sonar systems, wireless communications and seismology
\cite{Vantrees}. Traditional high-resolution methods for DOA
estimation such as the multiple signal classification (MUSIC) method
\cite{schimdt}, the root-MUSIC algorithm \cite{Barabell}, the
estimation of signal parameters via rotational invariance techniques
(ESPRIT) \cite{Roy} and subspace techniques
\cite{scharf,bar-ness,pados99,reed98,hua,goldstein,santos,qian,delamarespl07,xutsa,delamaretsp,kwak,xu&liu,delamareccm,wcccm,delamareelb,jidf,delamarecl,delamaresp,delamaretvt,jioel,delamarespl07,delamare_ccmmswf,jidf_echo,delamaretvt10,delamaretvt2011ST,delamare10,fa10,lei09,ccmavf,lei10,jio_ccm,ccmavf,stap_jio,zhaocheng,zhaocheng2,arh_eusipco,arh_taes,dfjio,rdrab,dcg_conf,dcg,dce,drr_conf,dta_conf1,dta_conf2,dta_ls,song,wljio,barc,jiomber,saalt},\cite{Steinwandt,Wang,Qiu}
exploit the eigenstructure of the input data matrix. These
techniques may fail for reduced data sets or low signal-to-noise
ratio (SNR) levels where the expected estimation error is not
asymptotic to the Cramér-Rao bound (CRB) \cite{Thomas}. The accuracy
of the estimates of the covariance matrix is of fundamental
importance in parameter estimation. Low levels of SNR or short data
records can result in significant divergences between the true and
the sample data covariance matrices. In practice, only a modest
number of data snapshots is available and when the number of
snapshots is similar to the number of sensor array elements, the
estimated and the true subspaces can differ significantly. Several
approaches have been developed with the aim of enhancing the
computation of the covariance matrix \cite{Carlson}-\cite{Qian} and
for dealing with large sensor-array systems large
\cite{mmimo,wence,Costa,delamare_ieeproc,TDS_clarke,TDS_2,switch_int,switch_mc,smce,TongW,jpais_iet,TARMO,badstbc,baplnc,keke1,kekecl,keke2,wlbd,Tomlinson,dopeg_cl,peg_bf_iswcs,gqcpeg,peg_bf_cl,Harashima,mbthpc,zuthp,rmbthp,Hochwald,BDVP},\cite{delamare_mber,rontogiannis,delamare_itic,stspadf,choi,stbcccm,FL11,jio_mimo,peng_twc,spa,spa2,jio_mimo,P.Li,jingjing,memd,did,bfidd,mbdf,bfidd,mserrr,shaowcl08}.

Diagonal loading \cite{Carlson} and shrinkage
\cite{Chen,ruan1,ruan2} techniques can enhance the estimate of the
data covariance matrix by weighing and individually increasing its
diagonal by a real constant. Nevertheless, the eigenvectors remain
the same, which leads to unaltered estimates of the signal and noise
projection matrices obtained from the enhanced covariance matrix.
Additionally, an improvement of the estimates of the covariance
matrix can be achieved by employing forward/backward averaging and
spatial smoothing approaches \cite{Pillai,Evans}. The former leads
to twice the number of the original samples and its corresponding
enhancement. The latter extracts the array covariance matrix as the
average of all covariance matrices from its sub-arrays, resulting in
a greater number of samples. Both techniques are employed in signal
decorrelation. An approach to improve MUSIC  dealing with the
condition in which the number of snapshots and the sensor elements
approach infinity was presented in \cite{Mestre}. Nevertheless, this
technique is not that effective for reduced number of snapshots.
Other approaches to deal with reduced data sets or low SNR levels
\cite{Gershman,Qian} consist of reiterating the procedure of adding
pseudo-noise to the observations which results in new estimates of
the covariance matrix. Then, the set of solutions is computed from
previously stored DOA estimates. In \cite{Vorobyov2}, two aspects
resulting from the computation of DOAs for reduced data sets or low
SNR levels have been studied using the root-MUSIC technique. The
first aspect dealt with the probability of estimated signal roots
taking a smaller magnitude than the estimated noise roots, which is
an anomaly that leads to wrong choices of the closest roots to the
unit circle. To mitigate this problem, different groups of roots are
considered as potential solutions for the signal sources and the
most likely one is selected \cite{Stoica}. The second aspect
previously mentioned, shown in \cite{Johnson}, refers to the fact
that a reduced part of the true signal eigenvectors exists in the
sample noise subspace (and vice-versa). Such coexistence has been
expressed by a Frobenius norm of the related irregularity matrix and
introduced its mathematical foundation. An iterative technique to
enhance the efficacy of root-MUSIC by reducing this anomaly making
use of the gradual reshaping of the sample data covariance matrix
has been reported. Inspired by the work in \cite{Vorobyov2}, we have
developed an ESPRIT-based method known as Two-Step KAI-ESPRIT
(TS-ESPRIT) \cite{Pinto}, which combines that modifications of the
sample data covariance matrix with the use of prior knowledge
\cite{Guerci1}-\cite{Guerci2} about the covariance matrix of a set
of impinging signals to enhance the estimation accuracy in the
finite sample size region. In practice, this prior knowledge could
be from the signals coming from known base stations or from static
users in a system. TS-ESPRIT determines the value of a correction
factor that reduces the undesirable terms in the estimation of the
signal and noise subspaces in an iterative process, resulting in
better estimates.

In this work \cite{Pinto2,Pinto3}, we present the Multi-Step KAI
ESPRIT (MS-KAI-ESPRIT) approach that refines the covariance matrix
of the input data via multiple steps of reduction of its undesirable
terms. This work presents the MS-KAI-ESPRIT in further detail, an
analysis of the mean squared error (MSE) of the data covariance
matrix free of undesired terms (side effects), a more accurate study
of the computational complexity and a comprehensive study of
MS-KAI-ESPRIT and other competing techniques for scenarios with both
uncorrelated and correlated signals. Unlike TS-ESPRIT, which makes
use of only one iteration and available known DOAs, MS-KAI-ESPRIT
employs multiple iterations and obtains prior knowledge on line. At
each iteration of MS-KAI-ESPRIT, the initial Vandermonde matrix is
updated by replacing an increasing number of steering vectors of
initial estimates with their corresponding refined versions. In
other words, at each iteration, the knowledge obtained on line is
updated, allowing the direction finding algorithm to correct the
sample covariance matrix estimate, which yields more accurate
estimates.

In summary, this work has the following contributions:
\begin{itemize}
    \item The proposed MS-KAI-ESPRIT technique.
    \item An MSE analysis of the covariance matrix obtained with the proposed MS-KAI-ESPRIT algorithm.
    \item A comprehensive performance study of MS-KAI-ESPRIT and competing techniques.
\end{itemize}

This paper is organized as follows. Section II describes the system
model. Section III presents the proposed MS-KAI-ESPRIT algorithm. In
section IV, an analytical study of the MSE of the data covariance
matrix free of side-effects is carried out together with a study of
the computational complexity of the proposed and competing
algorithms. In Section V, we present and discuss the simulation
results. Section VI concludes the paper. 

\section{System Model }
\label{sysmodel}

Let us assume that \textit{P} narrowband signals from far-field
sources impinge on a uniform linear array (ULA) of $M\ (M
> \textit{P})$ sensor elements from  directions
${\boldsymbol \theta}=[\theta_{1},\theta_{2},\ldots, \theta_P]^T$.
We also consider that the sensors are spaced from each other by a
distance $ d\leq\frac{\lambda_{c}}{2}$, where $\lambda_{c}$ is the
signal wavelength, and that without loss of generality, we have
${\frac{-\pi}{2}\leq\theta_{1}\leq\theta_{2}\leq\ldots
\leq\theta_P\leq \frac{\pi}{2}}$.

The $i$th data snapshot of the $M$-dimensional array output vector
can be modeled as
\begin{equation}
\bm x(i)=\bm A\,s(i)+\bm n(i),\qquad i=1,2,\ldots,N,
\label{model}
\end{equation}
where $\bm s(i)=[s_{1}(i),\ldots,s_{P}(i)]^T
\in\mathbb{C}^{\mathit{P\times1}}$ represents the zero-mean source
data vector, $\bm n(i) \in\mathbb{C}^{\mathit{M \times 1}}$ is the
vector of white circular complex Gaussian noise with zero mean and
variance $\sigma_n^2$, and $N$ denotes the number of available
snapshots.

The Vandermonde matrix $\bm A(\bm \Theta)=[\bm
a(\theta_{1}),\ldots,\bm a(\theta_{P})] \in\mathbb
{C}^{\mathit{M\times P}}$, known as the array manifold, contains the
array steering vectors $\bm a(\theta_j)$ corresponding to the $n$th
source, which can be expressed as
\begin{equation}
\bm a(\theta_n)=[1,e^{j2\pi\frac{d}{\lambda_{c}}
\sin\theta_n},\ldots,e^{j2\pi(M-1)\frac{d}{\lambda_{c}}\sin\theta_n}]^T,
\label{steer}
\end{equation}
where $n=1,\ldots, P$. Using the fact that $\bm s(i)$ and $\bm n(i)$
are modeled as uncorrelated linearly independent variables, the
$M\times M$ signal covariance matrix is calculated by
\begin{equation}
\bm R=\mathbb E\left[\bm x(i)\bm x^H(i)
\right]=\bm A\,\bm R_{ss}\bm A^H+
\sigma_n^2\bm I_M,
\label{covariance}
\end{equation}
where the superscript \textit{H} and $\mathbb E[\cdot]$ in $\bm
R_{ss}=\mathbb E[\bm s(i)\bm s^H(i)]$ and in $\mathbb E[\bm n(i)\bm
n^H(i)]=\sigma_n^2\bm I_M^{}$ denote the Hermitian transposition and
the expectation operator and $\bm I_M$ stands for the
$M$-dimensional identity matrix. Since the true signal covariance
matrix is unknown, it must be estimated and a widely-adopted
approach is the sample average formula given by
\begin{equation}
 \bm {\hat{R}}=\frac{1}{N} \sum\limits^{N}_{i=1}\bm x(i)\bm x^H(i),
\label{covsample}
\end{equation}
whose estimation accuracy is dependent on $N$.

\section{Proposed MS-KAI-ESPRIT Algorithm }

In this section, we present the proposed MS-KAI-ESPRIT algorithm and
detail its main features. We start by expanding \eqref{covsample}
using \eqref{model} as derived in \cite{Vorobyov2}:
\begin{eqnarray}
\bm {\hat{R}}=\frac{1}{N} \sum\limits^{N}_{i=1}(\bm A\,s(i)+\bm
n(i))\:(\bm A\,s(i)+\bm n(i))^H \nonumber\\= \bm
A\left\lbrace\frac{1}{N} \sum\limits^{N}_{i=1}\bm s(i)\bm
s^H(i)\right\rbrace\bm A^H+\:\frac{1}{N} \sum\limits^{N}_{i=1}\bm
n(i)\bm n^H(i)\;+\nonumber\\\underbrace{\bm A\left\lbrace\frac{1}{N}
    \sum\limits^{N}_{i=1}\bm s(i)\bm n^H(i)\right\rbrace\:
+\:\left\lbrace\frac{1}{N} \sum\limits^{N}_{i=1}\bm n(i)\bm
s^H(i)\right\rbrace\bm{A}^{H}}_{{"undesirable terms"}}
\label{expandedcovsample}
\end{eqnarray}
The first two terms of \text{$\bm {\hat{R}}$} in
\eqref{expandedcovsample} can be considered as estimates of the two
summands of \text{$\bm R$} given in  \eqref{covariance}, which
represent the signal and the noise components, respectively.  The
last two terms in \eqref{expandedcovsample} are undesirable side
effects, which  can be seen as estimates for the correlation between
the signal and the noise vectors. The system model under study is
based on noise vectors which are zero-mean  and also independent of
the signal vectors. Thus, the signal and noise components are
uncorrelated to each other. As a consequence, for a large enough
number of samples $N$, the last two  terms of
\eqref{expandedcovsample} tend to zero. Nevertheless, in practice
the number of available samples can be limited. In such situations,
the last two terms in \eqref{expandedcovsample} may have significant
values, which causes the deviation of the estimates of the signal
and the noise subspaces from the true signal and noise subspaces.

The key point of the proposed MS-KAI-ESPRIT algorithm is to modify
the sample data covariance matrix estimate at each iteration by
gradually incorporating the knowledge provided by the newer
Vandermonde matrices which progressively embody the refined
estimates from the preceding iteration. Based on these updated
Vandermonde matrices, refined estimates of the projection matrices
of the signal and noise subspaces are calculated. These estimates of
projection matrices associated with the initial sample covariance
matrix estimate and the reliability factor are employed to reduce
its side effects and allow the algorithm to choose the set of
estimates that has the highest likelihood of being the set of the
true DOAs. The modified covariance matrix is computed by computing a
scaled version of the undesirable terms of $\bm {\hat{R}}$, as
pointed out in \eqref{expandedcovsample}.

The steps of the proposed algorithm are listed in Table
\ref{Multi_Step_KAI}. The algorithm starts by computing the sample
data covariance matrix \eqref{covsample}. Next, the DOAs are
estimated using the ESPRIT algorithm. The superscript
$(\cdot)^{(1)}$ refers to the estimation task performed in the first
step. Now, a procedure consisting of $n=1:P$ iterations starts by
forming the Vandermonde matrix using the DOA estimates. Then, the
amplitudes of the sources are estimated such that the square norm of
the differences between the observation vector and the vector
containing estimates and the available known DOAs is minimized. This
problem can be formulated \cite{Vorobyov2} as:
\begin{eqnarray}
    \hat{\bm{s}}(i)=\arg\min_{\substack{\bm
    s}}\parallel\bm{x}(i)-\hat{\bm{A}}\mathbf{s}\parallel^2_2.
    \label{minimization1}
\end{eqnarray}
The minimization of \eqref{minimization1} is achieved using the
least squares technique and the solution is described by
\begin{equation}
\hat{\bm{s}}(i)=(\mathbf{\hat{A}}^{H}\:\mathbf{\hat{A}})^{-1}\:\mathbf{\hat{A}}\:\bm{x}(i)
\label{minimization2}
\end{equation}
The noise component is then estimated as the difference between the
estimated signal and the observations made by the array, as given by
\begin{eqnarray}
 \hat{\bm n}(i)=\bm x(i)\:-\: \hat{\bm A}\:\hat{\bm s}(i).
\label{noise_component}
\end{eqnarray}
After estimating  the signal and noise vectors, the third term in
\eqref{expandedcovsample} can be computed as:
\begin{align}
\bm{V}&\triangleq \hat{\bm{A}}\left\lbrace\frac{1}{N}
\sum\limits^{N}_{i=1}\bm \hat{\mathbf{s}}(i)\bm
\hat{\mathbf{n}}^H(i)\right\rbrace\nonumber\\&=\hat{\bm{A}}\left\lbrace\frac{1}{N}
\sum\limits^{N}_{i=1}(\mathbf{\hat{A}}^{H}\:\mathbf{\hat{A}})^{-1}\mathbf{\hat{A}}^{H}\bm{x}(i)\right.\nonumber\\&\left.\times(\bm{x}^{H}(i)-\bm{x}^{H}(i)\hat{\mathbf{A}}(\hat{\mathbf{A}}^{H}\hat{\mathbf{A}})^{-1}\:\hat{\mathbf{A}}^{H})\right\rbrace\nonumber\\&=\mathbf{\hat{Q}}_{A}\left\lbrace\frac{1}{N}
\sum\limits^{N}_{i=1}
\bm{x}(i)\bm{x}^H(i)\:\left(\mathbf{I}_{M}\:-\:\hat{\mathbf{Q}}_{A}\right)
\right\rbrace\nonumber\\&=\mathbf{\hat{Q}}_{A}\:\mathbf{\hat{R}}\:\mathbf{\hat{Q}}_{A}^{\perp},
\label{terms_deducted}
\end{align}
where
\begin{equation}
\mathbf{\hat{Q}}_{A}\triangleq \mathbf{\hat{A}}\:(\mathbf{\hat{A}}^{H}\:\mathbf{\hat{A}})^{-1}\:\mathbf{\hat{A}}^{H}
\label{signal_projection}
\end{equation}
is an estimate of the projection matrix of the signal subspace, and
\begin{equation}
\mathbf{\hat{Q}}_{A}^{\perp}\triangleq\mathbf{I}_{M}\:-\:\mathbf{\hat{Q}}_{A}
\label{noise_projection}
\end{equation}
is an estimate of the projection matrix of the noise subspace.

Next, as part of the procedure consisting of $n=1:P $ iterations,
the modified data covariance matrix $\mathbf{\hat{R}}^{(n+1)}$ is
obtained by computing a scaled version of the estimated terms from
the initial sample data covariance matrix as given by
\begin{equation}
\label{modified_data_covariance}
\mathbf{\hat{R}}^{(n+1)} = \mathbf{\hat{R}}\:-\:\mathrm{\mu}\:(\mathbf{V}^{(n)}\:+\:\mathbf{V}^{(n)H}),
\end{equation}
where the superscript $(\cdot)^{(n)}$ refers to the $n^{th} $
iteration performed. The scaling or reliability factor \text{$\mu $}
increases from 0 to 1 incrementally, resulting in modified data
covariance matrices. Each of them gives origin to new estimated DOAs
also denoted by the superscript  $(\cdot)^{(n+1)}$ by using the
ESPRIT algorithm, as briefly described ahead.


In this work, the rank \textit{P} is assumed to be known, which is
an assumption frequently found in the literature. Alternatively, the
rank \textit{P} could be estimated by model-order selection schemes
such as Akaike´s Information Theoretic Criterion (AIC) \cite{Schell}
and the Minimum Descriptive Length (MDL) Criterion \cite{Rissanen}.

In order to estimate the signal and the orthogonal subspaces from
the data records, we may consider two approaches
\cite{Vaccaro,Haardt}: the direct data approach and the covariance
approach. The direct data approach makes use of singular value
decomposition(SVD) of the data matrix $\mathbf{X}$, composed of the
$i$th data snapshot \eqref{model} of the $M$-dimensional array data
vector:
\begin{align}
\bm X=&[\bm
{x}(1),\bm{x}(2),\ldots,\bm{x}(N)]\nonumber\\=&\bm A[\bm
{s}(1),\bm{s}(2),\ldots,\bm{s}(N)]+[\bm
{n}(1),\bm
{n}(2),\ldots,\bm{n}(N)]\nonumber\\=&\bm{A(\Theta)\;S}+\:\bm{N}\;\in\mathbb{C}^{\mathit{M \times N}}
\label{Data_matrix}
\end{align}

Since the number of the sources is assumed known or can be estimated
by AIC\cite{Schell} or MDL\cite{Rissanen} , as previously mentioned,
we can write $\mathbf{X}$ as:

\begin{align}
\begin{array}{ccc}
\bm{X}
\end{array}& =\left[ \begin{array}{ccc}
\mathbf{\hat{U}}_{s} & \mathbf{\hat{U}}_{n}  \\
\end{array} \right]\left[ \begin{array}{ccc}
\mathbf{\hat{\Gamma}}_{s} & 0  \\
0 & \mathbf{\hat{\Gamma}}_{n}
\end{array} \right]\left[ \begin{array}{ccc}
\mathbf{\hat{U}}_{s}^{H} \\
\mathbf{\hat{U}}_{n}^{H}
\end{array} \right],
\label{SVD}
\end{align}
where the diagonal matrices $\mathbf{\hat{\Gamma}}_{s}$ and
$\mathbf{\hat{\Gamma}}_{n}$ contain the $\mathit{P}$ largest
singular values and the $\mathit{M-P}$ smallest singular values,
respectively. The estimated signal subspace $\mathbf{\hat{U}}_{s}$
$\in\mathbb{C}^{\mathit{M \times P}}$ consists of the singular
vectors corresponding to $\mathbf{\hat{\Gamma}}_{s}$ and the
orthogonal subspace $\mathbf{\hat{U}}_{n}$ $\in\mathbb{C}^{\mathit{M
\times (M-P)}}$ is related to $\mathbf{\hat{\Gamma}}_{n}$. If the
signal subspace is estimated a rank-\textit{P} approximation of the
SVD can be applied.

The covariance approach applies the eigenvalue decomposition (EVD)
of the sample covariance matrix \eqref{covsample}, which is related
to the data matrix \eqref{Data_matrix}:
    \begin{equation}
    \bm {\hat{R}}=\frac{1}{N} \sum\limits^{N}_{i=1}\bm x(i)\bm x^H(i)=\frac{1}{N}\bm {X}\bm {X}^H\;\in\mathbb{C}^{\mathit{M \times M}
        \label{rel_datamatrix_covsample}}
    \end{equation}
Then, the EVD of \eqref{rel_datamatrix_covsample} can be carried out
as follows:
\begin{equation}
\begin{array}{ccc}
\mathbf{\hat{R}}
\end{array} =\left[ \begin{array}{ccc}
\mathbf{\hat{U}}_{s} & \mathbf{\hat{U}}_{n}  \\
\end{array} \right]\left[ \begin{array}{ccc}
\mathbf{\hat{\Lambda}}_{s} & 0  \\
0 & \mathbf{\hat{\Lambda}}_{n}
\end{array} \right]\left[ \begin{array}{ccc}
\mathbf{\hat{U}}_{s}^{H} \\
\mathbf{\hat{U}}_{n}^{H}
\end{array} \right],
\label{Detalha_ESPRIT}
\end{equation}
where the diagonal matrices $\mathbf{\hat{\Lambda}_{s}}$ and
$\mathbf{\hat{\Lambda}_{n}}$ contain the \textit{P} largest and the
    \textit{M-P} smallest eigenvalues, respectively.
    The estimated signal subspace $\mathbf{\hat{U}}_{s}$ $\in$
    $\mathbb{C}^{\mathit{M\times P}}$ corresponding to $\mathbf{\hat{\Gamma}_{s}}$
    and the orthogonal subspace $\mathbf{\hat{U}}_{n}$ $\in$
    $\mathbb{C}^{\mathit{M\times(M-P) }}$ complies with $\hat{\Gamma}_{n}$.
    If the signal subspace is estimated a rank-P approximation of the EVD can be applied.
With infinite precision arithmetic, both SVD and EVD can be
considered equivalent. However, as in practice, finite precision
arithmetic is employed, 'squaring' the data to obtain the Gramian
$\bm {X}\bm {X}^H$ \eqref{rel_datamatrix_covsample} can result in
round-off errors and overflow. These are potential problems to be
aware when using the covariance approach. 

Now, we can briefly review ESPRIT. We start by forming a twofold
subarray configuration, as each row of the array steering matrix
$\bm A(\bm \Theta)$ corresponds to one sensor element of the antenna
array. The subarrays are specified by two $\mathit{(s\times
M)}$-dimensional selection matrices $\mathbf{J_{1}}$  and
$\mathbf{J_{2}}$ which choose $\mathit{s}$ elements of the
$\mathit{M}$ existing sensors, respectively, where $\mathit{s}$ is
in the range  $\mathit{P\leq s < M}$. For maximum overlap, the
matrix $\mathbf{J_{1}}$ selects  the first $\mathit{s=M-1}$ elements
and the matrix $\mathbf{J_{2}}$ selects the last $\mathit{s=M-1}$
rows of $\bm A(\bm \Theta)$.

Since the matrices $\mathbf{J_{1}}$ and $\mathbf{J_{2}}$ have now
been computed, we can estimate the  operator $ \mathbf{\Psi} $ by
solving the approximation of the shift invariance equation
\eqref{shift_invariance_equation} given by
\begin{equation}
\mathbf{J}_{1}\:\mathbf{\hat{U}}_{s}\:\mathbf{\Psi}\:\approx\:\mathbf{J}_{2}\:\mathbf{\hat{U}}_{s}.
\label{shift_invariance_equation}
\end{equation}
where $\hat{U}_{s}$ is obtained in \eqref{Detalha_ESPRIT}.

Using  the least squares (LS) method, which yields
\begin{equation}
\hat{\mathbf{\Psi}}=\arg\min_{\substack{\mathbf
{\Psi}}}\parallel\mathbf{J}_{2}\:\mathbf{\hat{U}}_{s}\:-\:\mathbf{J}_{1}\:\mathbf{\hat{U}}_{s}\:\mathbf{\Psi}\parallel_{F}\:=\:\left(
\mathbf{J}_{1}\:\hat{\mathbf{U}}_{s}\right)
^{\dagger}\:\mathbf{J}_{2}\:\hat{\mathbf{U}}_{s},
\end{equation}
where $\parallel\cdot\parallel_{F}$ denotes the Frobenius norm and
$\left( \cdot\right)^{\dagger}$ stands for the pseudo-inverse.

Lastly, the eigenvalues $\lambda_{i}$ of $\hat{\mathbf{\Psi}}$
contain the estimates of the spatial frequencies $\gamma_{i}$
computed as:
\begin{equation}
\gamma_{i}\:=\:\arg\left(\lambda_{i} \right),
\label{spatial_frequencies}
\end{equation}
so that the DOAs can be calculated as:
\begin{equation}
\hat{\theta}_{i}\:=\:\arcsin\left(\frac{\gamma_{i}\:\lambda_{c}}{2\pi\:\mathit{d}} \right)
\label{doas_ESPRIT}
\end{equation}
where for \eqref{spatial_frequencies} and \eqref{doas_ESPRIT}
$\mathrm{i=1,\cdots,P}$.

Then, a new Vandermonde matrix $\mathbf{\hat{B}}^{(n+1)}$ is formed
by the steering vectors of those refined estimates of the DOAs. By
using this updated matrix, it is possible to compute the refined
estimates of the projection matrices of the signal \text{$
\mathbf{\hat{Q}}_{B}^{(n+1)} $} and the noise \text{$
\mathbf{\hat{Q}}_{B}^{(n+1)\perp} $} subspaces.

Next, employing the refined estimates of the projection matrices,
the initial sample data matrix, $\bm {\hat{R}}$, and the number of
sensors and sources, the stochastic maximum likelihood objective
function $\mathit{U^{(n+1)}(\mu)}$ \cite{Stoica} is computed for
each value of \text{$\mu $} at the $n^{th}$ iteration, as follows:
\begin{equation}
\begin{split}
\mathit{U^{(n+1)}(\mu)} & =\mathrm{ln\:det}
\Big(\mathbf{\hat{Q}}_{B}^{(n+1)}\:\mathbf{\hat{R}}\:\mathbf{\hat{Q}}_{B}^{(n+1)}
\\
&  \quad + \dfrac{{\rm
Trace}\{\mathbf{\hat{Q}}_{B}^{\perp\:(n+1)}\:\mathbf{\hat{R}}\}}
{\mathrm{M-P}}\:\mathbf{\hat{Q}}_{B}^{\:(n+1)\perp} \Big).
\end{split}
\label{SML_objective_function}
\end{equation}

The previous computation selects the set of unavailable DOA
estimates that have a higher likelihood at each iteration. Then, the
set of estimated DOAs corresponding to the optimum value of
\text{$\mu $} that minimizes \eqref{SML_objective_function} also at
each $n^{th}$ iteration is determined. Finally, the output of the
proposed MS-KAI-ESPRIT algorithm is formed by the set of the estimates
obtained at the $P^{th}$ iteration, as described in Table
\ref{Multi_Step_KAI}.

\begin{table}[htb!]
    \centering
    \small
    \caption{Proposed MS-KAI-ESPRIT Algorithm}\smallskip
    \scalebox{1.0}\medskip{
        \begin{tabular}{|r l|}
            \hline

            \multicolumn{2}{|l|}{\small $\textbf{\underline{Inputs}:}$}\\[0.7ex]
            \multicolumn{2}{|l|}{\small$\mathit{M}$,\hspace{2mm}$\mathit{d}$,\hspace{2mm}$\lambda$,\hspace{2mm}$\mathit{N}$,\hspace{2mm}$\mathit{P}$ }\\[0.6ex]
            \multicolumn{2}{|l|}{\small\text{Received vectors}  $\bm x(1)$,\hspace{2mm}$\bm x(2)$,$\cdots$, $\bm x(N)$}\\[0.6ex]

            \multicolumn{2}{|l|}{\small $\textbf{\underline{Outputs}:}$}\\[0.6ex]

            \multicolumn{2}{|l|}{\small\text{Estimates}\hspace{1mm}$\mathit{\hat{\theta}_{1}^{(n+1)}(\mu\,opt)}$,\hspace{2mm}$\mathit{\hat{\theta}_{2}^{(n+1)}(\mu\,opt)}$,$\cdots$,\hspace{2mm}$\mathit{\hat{\theta}_{P}^{(n+1)}(\mu\,opt)}$} \\[3.1ex]
            \hline

            \multicolumn{2}{|l|}{\small $\textbf{\underline{First step}:}$}\\[0.9ex]

            \multicolumn{2}{|l|}{\small $\mathbf{\hat{R}}=\frac{1}{N} \sum\limits^{N}_{i=1}\bm x(i)\bm x^H(i)$}\\[1.8ex]

            \multicolumn{2}{|l|}{\small $\{\mathit{\hat{\theta}_{1}}^{(1)},\:\mathit{\hat{\theta}_{2}}^{(1)},\cdots,\mathit{\hat{\theta}_{P}}^{(1)}\}\;\;\underleftarrow{ESPRIT}$ $\:(\mathbf{\hat{R}},P,d,\lambda)$}\\[1.2ex]

            \multicolumn{2}{|l|}{\small$\mathbf{\hat{A}}^{(1)}=\left[\mathbf{a}(\mathit{\hat{\theta}_{1}^{(1)}}),\mathbf{a}(\mathit{\hat{\theta}_{2}^{(1)}}),\cdots,\mathbf{a}(\mathit{\hat{\theta}_{P}^{(1)}})\right]$} \\ [1.1ex]

            \multicolumn{2}{|l|}{\small $\textbf{\underline{Second step}:}$}\\[0.9ex]

            \multicolumn{2}{|l|}{\small \textbf{for}  \textit{n}\hspace{1mm}=\hspace{2mm}\text{1}\hspace{1mm}:\hspace{1mm}\textit{P}}\\[0.6ex]

            \multicolumn{2}{|l|}{\small $\mathbf{\hat{Q}}_{A}^{(n)}= \mathbf{\hat{A}}^{(n)}\:(\mathbf{\hat{A}}^{(n)H}\:\mathbf{\hat{A}}^{(n)})^{-1}\:\mathbf{\hat{A}}^{(n)H}$ }\\[0.9ex]

            \multicolumn{2}{|l|}{\small$\mathbf{\hat{Q}}_{A}^{(n)\perp}=\mathbf{I}_{M}\:-\:\mathbf{\hat{Q}}_{A}^{(n)}$ }\\[1.0ex]

            \multicolumn{2}{|l|}{\small $\mathbf{V}^{(n)}=\mathbf{\hat{Q}}_{A}^{(n)}\:\mathbf{\hat{R}}\:\mathbf{\hat{Q}}_{A}^{(n)\perp}$}\\[1.8ex]

            \multicolumn{2}{|l|}{\small \textbf{for} $\mathbf{\mu=}\hspace{1mm}\text{0}:\text{$\iota$ \hspace{1mm}:\hspace{1mm}1} $}\\[0.9ex]

            \multicolumn{2}{|l|}{\small $ \mathbf{\hat{R}}^{(n+1)} = \mathbf{\hat{R}}\:-\:\mathrm{\mu}\:(\mathbf{V}^{(n)}\:+\:\mathbf{V}^{(n)H})$} \\[0.9ex]

            \multicolumn{2}{|l|}{\small $\{\mathit{\hat{\theta}_{1}}^{(n+1)},\:\mathit{\hat{\theta}_{2}}^{(n+1)},\cdots,\mathit{\hat{\theta}_{P}}^{(n+1)}\}\;\;\underleftarrow{ESPRIT}$ $\:(\mathbf{\hat{R}}^{(n+1)},\:P,d,\lambda)$}\\[1.4ex]

            \multicolumn{2}{|l|}{\small$\mathbf{\hat{B}}^{(n+1)}=\left[\mathbf{a}(\mathit{\hat{\theta}_{1}^{(n+1)}}),\mathbf{a}(\mathit{\hat{\theta}_{2}^{(n+1)}}),\cdots,\mathbf{a}(\mathit{\hat{\theta}_{P}^{(n+1)}})\right]$} \\ [1.4ex]

            \multicolumn{2}{|l|}{\small $\mathbf{\hat{Q}}_{B}^{(n+1)}= \mathbf{\hat{B}}^{(n+1)}\:(\mathbf{\hat{B}}^{(n+1)H}\:\mathbf{\hat{B}}^{(n+1)})^{-1}\:\mathbf{\hat{B}}^{(n+1)H}$}\\[1.2ex]

            \multicolumn{2}{|l|}{\small$\mathbf{\hat{Q}}_{B}^{(n+1)\perp}=\mathbf{I}_{M}\:-\:\mathbf{\hat{Q}}_{B}^{(n+1)}$}\\[1.2ex]

            \multicolumn{2}{|l|}{\small $\mathit{U^{(n+1)}(\mu)}=\mathrm{ln\:det}\left(\cdot\right),$}\\[1.1ex]

            \multicolumn{2}{|l|}{\small$\left(\cdot\right) =\left(\mathbf{\hat{Q}}_{B}^{(n+1)}\:\mathbf{\hat{R}}\:\mathbf{\hat{Q}}_{B}^{(n+1)}+\dfrac{{\rm Trace}\{\mathbf{\hat{Q}}_{B}^{\perp\:(n+1)}\:\mathbf{\hat{R}}\}} {\mathrm{M-P}}\:\mathbf{\hat{Q}}_{B}^{\:(n+1)\perp}\right)$}\\[1.1ex]

            \multicolumn{2}{|l|}{\small $\mathit{\mu}_{\mathrm{opt}}^{(n+1)}=\arg \min \hspace{1mm}\mathit{U^{(n+1)}(\mu)}$}\\[1.0ex]

            \multicolumn{2}{|l|}{\small $\mathrm{DOAs}^{(n+1)}= \left(\ast\right),$}\\[1.1ex]

            \multicolumn{2}{|l|}{\small$\left(\ast\right) =
                 \{\mathit{\hat{\theta}_{1}^{(n+1)}(\mu\,opt)}$,\hspace{2mm}$\mathit{\hat{\theta}_{2}^{(n+1)}(\mu\,opt)}$,$\cdots$,\hspace{2mm}$\mathit{\hat{\theta}_{P}^{(n+1)}(\mu\,opt)}\}$}\\[1.0ex]

            \multicolumn{2}{|l|}{\small $\mathbf{\hat{A}}^{(n+1)}=\left\{\mathbf{a}(\mathit{\hat{\theta}_{\{1,\cdots,n\}}^{(n+1)}})\right\}\bigcup\left\{\mathbf{a}(\mathit{\hat{\theta}_{\{1,\cdots,P\}\,-\,\{1,\cdots,n\}}^{(1)}})\right\}$}\\[1.0ex]

            \multicolumn{2}{|l|}{\small \textbf{end for}}\\[1.0ex]

            \multicolumn{2}{|l|}{\small \textbf{end for}}\\[1.0ex]

            \hline
        \end{tabular}
    }
    \label{Multi_Step_KAI}

\end{table}
\section{ Analysis}
\label{Analysis}

In this section, we carry out an analysis of the MSE of the data
covariance matrix free of side effects along with a study of the
computational complexity of the proposed MS-KAI-ESPRIT and existing
direction finding algorithms.

\subsection{MSE Analysis}
\label{MSE_analysis}

In this subsection we show that at the first of the $P$ iterations,
the MSE of the data covariance matrix free of side effects
$\mathbf{\hat{R}}^{(n+1)}$  is less than or equal to the MSE of that
of the original one $\mathbf{\hat{R}}$. This can be formulated as:
\begin{align}
\mathrm{MSE}\left( \mathbf{\hat{R}}^{(n+1)}\right) \leq\mathrm{MSE}\left(\mathbf{\hat{R}}\right)
\label{problem_formulation1}
\end{align}
or, alternatively, as
\begin{align}
\mathrm{MSE}\left( \mathbf{\hat{R}}^{(n+1)}\right)- \mathrm{MSE}\left(\mathbf{\hat{R}}\right) \leq0
\label{problem_formulation2}
\end{align}
The proof of this inequality is provided in the Appendix.

\subsection{Computational Complexity Analysis}
\label{computational_analysis}

In this section, we evaluate the computational cost of the proposed
MS-KAI-ESPRIT algorithm which is compared to the following classical
subspace methods: ESPRIT \cite{Roy}, MUSIC \cite{schimdt},
Root-MUSIC \cite{Barabell}, Conjugate Gradient (CG) \cite{Semira},
Auxiliary Vector Filtering (AVF) \cite{Grover} and TS-ESPRIT
\cite{Pinto}. The ESPRIT and MUSIC-based methods use the Singular
Value Decomposition (SVD) of the sample covariance matrix
\eqref{covsample}. The computational complexity of MS-KAI-ESPRIT  in
terms of number of multiplications and additions is depicted in
Table \ref{Comput_Complexity1}, where $\mathrm{\tau}=\frac{1}{ \iota} +1$. The increment
${ \iota}$ is defined in Table \ref{Multi_Step_KAI}. As can be seen,
for this specific configuration used in the simulations
\ref{simulations} MS-KAI-ESPRIT shows a relatively high
computational burden with
$\mathcal{O}\mathit{(P\tau(3M^{3}+8MN^{2}))}$, where $\tau$ is
typically an integer that ranges from $ 1$  to $20 $. It can be
noticed that for the configuration used in the simulations $(P=4,
M=40, N=25)$ $ 3M^{3}$ and $8MN^{2}$ are comparable, resulting in
two dominant terms. It can also be seen that the number of multiplications required by the proposed algorithm is more significant than the number of additions. For this reason, in Table \ref{Comput_Complexity2},  we computed only the  computational burden of the previously mentioned algorithms in terms of multiplications for the purpose of comparisons. In that table, $\Delta$  stands for the search step.

\begin{figure}[!h]

    \centering 
    \includegraphics[width=8cm,height=6cm]{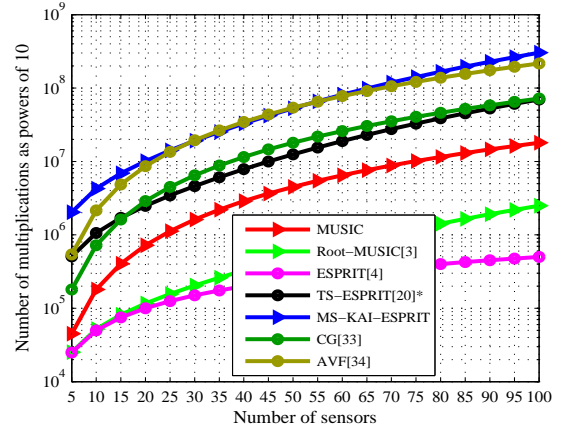} 
    \vspace{-1.0em}\caption{Number of multiplications as powers of 10 versus number of sensors  for $P=4$, $N=25$. }

    \label{figura:Multiplications_reviewer2}  
\end{figure}

Next, we will evaluate the influence of the number of sensor
elements on the number of multiplications based on the specific
configuration described in Table \ref{sysmodel}. Supposing
$\mathrm{P=4}$ narrowband signals impinging a ULA of $\mathrm{M}$
sensor elements and $\mathrm{N=25} $ available snapshots, we obtain
Fig. \ref{figura:Multiplications_reviewer2}. We can see the main trends in
terms of computational cost measured in multiplications of the
proposed and analyzed algorithms.
By examining Fig. \ref{figura:Multiplications_reviewer2}, it can be noticed
that in the range $M=\left[ 20\ 70\right] $ sensors, the curves
describing the exact number of multiplications in MS-KAI-ESPRIT and
AVF tend to merge. 
For $M=40$, this ratio tends to $1$, i.e. the number
of multiplications are almost equivalent.
\begin{table}[!h]
    \caption{Computational complexity - MS-KAI-ESPRIT}
    \vspace{2mm}
    \centering
    \begin{tabular}{|l|l|p{6cm}|}
        \hline
        & \underline{Multiplications} \\

        \rule{0pt}{3ex}  & $P\:\mathrm{\tau [\frac{10}{3}M^{3}+M^{2}(3P+2)+M(\frac{5}{2}P^{2}+\frac{1}{2}P+8N^{2})}$  \\[1pt]

        MS-KAI&$\mathrm{+P^{2}(\frac{17}{2}P+\frac{1}{2})]}$ \\[1pt]
        -ESPRIT& \\[1pt]
        (Proposed)& $\mathrm{+P\:[2M^{3}+M^{2}(P)+M(\frac{3}{2}P^{2}+\frac{1}{2}P)+P^{2}(\frac{P}{2}+\frac{3}{2}) ]}$ \\[1pt]

        & $\mathrm{+2M^{2}(P)+M(P^{2}-P+8N^{2})+P^{2}(8P-1)}$\\[2pt]

        & \underline{Additions} \\[2pt]

        \rule{0pt}{3ex}  & $P\:\mathrm{\tau
            [\frac{10}{3}M^{3}+M^{2}(3P-1)+M(\frac{5}{2}P^{2}-\frac{9}{2}P+8N^{2})}$  \\[1pt]

        &$\mathrm{+P(8P^{2}-2P-\frac{5}{2})]}$ \\[1pt]

        & $\mathrm{+P\:[2M^{3}+M^{2}(P-2)+M(\frac{3}{2}P^{2}-\frac{1}{2}P)-P(P+\frac{1}{2}) ]}$ \\[1pt]

        & $\mathrm{+2M^{2}(P)+M(P^{2}-4P+8N^{2})+P(8P^{2}-P-2)}$\\[2pt]
        \hline

    \end{tabular}

    \label{Comput_Complexity1}
\end{table}
\begin{table}[!h]
    \caption{Computational complexity - other algorithms}
    \vspace{2mm}
    \centering
    \begin{tabular}{|l|l|p{6cm}|}
        \hline
        Algorithm & Multiplications \\[2pt]
        \hline
        MUSIC \cite{schimdt} & $\mathrm{\frac{180}{\Delta}[M^{2}+M(2-P)-P]+8MN^{2}}$  \\[2pt]
        \hline
        root-MUSIC\cite{Barabell} & $\mathrm{2M^{3}-M^{2}P+8MN^{2}}$  \\[2pt]
        \hline
        AVF \cite{Grover} & $\mathrm{\frac{180}{\Delta}[M^{2}(3P+1)+M(4P-2)+P+2]}$\\[1pt]
        &$\mathrm{+M^{2}N}$  \\[2pt]
        \hline
        CG \cite{Semira} & $\mathrm{\frac{180}{\Delta}[M^{2}(P+1)+M(6P+2)+P+1]+M^{2}N}$  \\[2pt]
        \hline
        ESPRIT\cite{Roy} & $\mathrm{2M^{2}P+M(P^{2}-2P+8N^{2})+8P^{3}-P^{2} }$  \\[2pt]
        \hline
        & $\mathrm{\tau[3M^{3}+M^{2}(3P+2)+M(\frac{5}{2}P^{2}-\frac{3}{2}P+8N^{2})}$  \\[1pt]

        &$\mathrm{+P^{2}(\frac{17}{2}P+\frac{1}{2})+1]}$ \\[1pt]

        TS-ESPRIT \cite{Pinto}*& $\mathrm{+[2M^{3}+M^{2}(3P)+M(\frac{5}{2}P^{2}-\frac{3}{2}P+8N^{2})}$ \\[1pt]

        & $\mathrm{+P^{2}(\frac{17}{2}P+\frac{1}{2})]}$\\[2pt]

        \hline
    \end{tabular}

    \label{Comput_Complexity2}
\end{table}

\section{Simulations}
\label{simulations}

In this section, we examine the performance  of the proposed
MS-KAI-ESPRIT in terms of probability of resolution and RMSE and
compare them to the standard ESPRIT \cite{Roy}, the Iterative ESPRIT
(IESPRIT), which is also developed here by combining the approach in
\cite{Vorobyov2} that exploits knowledge of the structure of the
covariance matrix and its perturbation terms, the Conjugate Gradient
(CG) \cite{Semira}, the Root-MUSIC \cite{Barabell}, and the MUSIC
\cite{schimdt} algorithms. Despite TS-ESPRIT is based on the
knowledge of available known DOAS and the proposed MS-KAI-ESPRIT
does not have access to prior knowledge, TS-ESPRIT is plotted with
the aim of illustrating the comparisons. For a fair comparison in
terms of RMSE and probability of resolution of all studied
algorithms, we suppose that we do not have prior knowledge, that is
to say that although we have available known DOAs, we compute
TS-ESPRIT as they were unavailable. We employ a ULA with \textit{
M=40} sensors, inter-element spacing $\Delta=\frac{\lambda_{c}}{2}$
and assume there are four uncorrelated complex Gaussian signals with
equal power impinging on the array. The closely-spaced sources are
separated by \textit{$2.4^{o}$}, at $\mathrm
(10.2^{o},12.6^{o},15^{o},17.4^{o})$,  and the number of available
snapshots is \textit{N}=25. For TS-ESPRIT, as previously mentioned,
we presume a priori knowledge of the last true DOAS
$\mathrm(15^{o},17.4^{o})$

\begin{figure}[!h]
    \centering 
    \includegraphics[width=8cm,height=6cm]{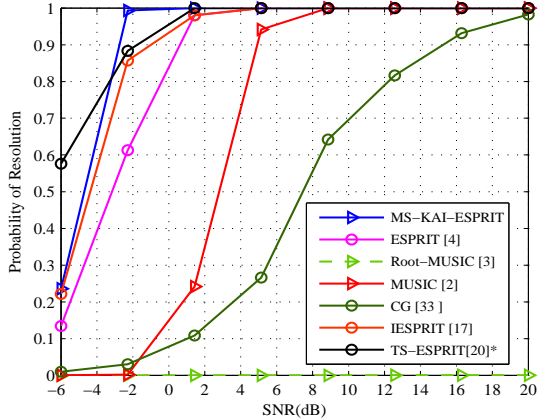} 
    \vspace{-1.0em}\caption{Probability of resolution versus SNR with $P=4$ uncorrelated sources, $M=40$, $N=25$, $L=100$ runs}
    \label{figura:DSP_PR_2ponto4deg_40sens_25snap_100runs}
\end{figure}

In Fig. \ref{figura:DSP_PR_2ponto4deg_40sens_25snap_100runs}, we
show the probability of resolution versus SNR. We take into account
the criterion \cite{Stoica3}, in which two sources with DOA
$\theta_{1}$ and $\theta_{2}$  are said to be resolved if their
respective estimates $\hat{\theta}_{1}$ and $\hat{\theta}_{2}$ are
such that both $\left|\hat{\theta}_{1} -\theta_{1}\right|$ and
$\left|\hat{\theta}_{2} -\theta_{2}\right|$ are less than
$\left|\theta_{1} -\theta_{2}\right|/2$. The proposed MS-KAI-ESPRIT
algorithm outperforms IESPRIT developed here, based on
\cite{Vorobyov2}, and the standard ESPRIT \cite{Roy} in the range
between $-6$ and $ 5dB $ and MUSIC \cite{schimdt} from  $-6$ to $
8.5dB $. MS-KAI-ESPRIT also outperforms CG \cite{Semira} and Root-Music
\cite{Barabell} throughout the whole range of values. The poor
performance of the latter could be expected from the results for two
closed signals obtained in \cite{Vorobyov2}. When compared to
TS-ESPRIT, which as previously discussed, was supposed to have the
best performance, the proposed MS-KAI-ESPRIT algorithm is outperformed
by the former only in the range between $ -6 $ and $ -2dB $. From
this last point to $ 20dB $ its performance is superior or equal to
the other algorithms.

\begin{figure}[!h]
    \centering 
    \includegraphics[width=8cm, height=6cm]{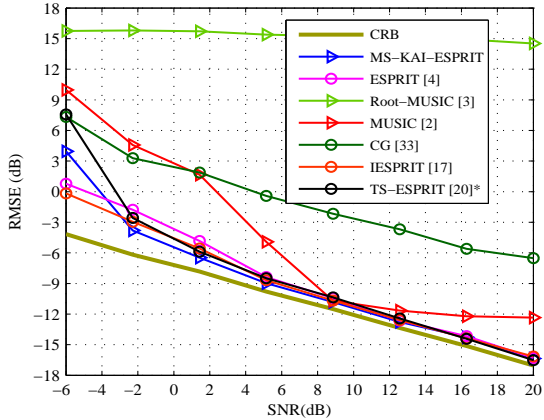} 
    \vspace{-1.0em}\caption{RMSE and the square root of CRB versus SNR with $P=4$ uncorrelated sources, $M=40$, $N=25$, $L=100$ runs}
    \label{figura:DSP_RMSE_CRB_2ponto4deg_40sens_25snap_100runs}
\end{figure}

In Fig. \ref{figura:DSP_RMSE_CRB_2ponto4deg_40sens_25snap_100runs},
it is shown the RMSE in dB  versus SNR, where  the term CRB refers
to the square root of the deterministic Cramér-Rao bound
\cite{Stoica4}. The RMSE is defined as:
\begin{equation}
\centering \mathrm{RMSE}
=\sqrt{\frac{1}{L\:P}\sum\limits^{L}_{l=1}\sum\limits^{P}_{p=1}\bm(\theta_p
-\bm \hat{\theta}_p(l))^{2}}, \label{RMSE_run}
\end{equation}
where $L$ is the number of trials.

The results show the superior performance of MS-KAI-ESPRIT in the
range between $-2.5$ and $5$ dB. From this last point to $20$ dB,
MS-KAI-ESPRIT, IESPRIT, ESPRIT and TS-ESPRIT have similar
performance. The only range in which MS-KAI-ESPRIT is outperformed
lies in the range between $-6$ and $-2.5$ dB. From this last point
to $20$ dB its performance is better or similar to the others.

Now, we focus on the performance of MS-KAI-ESPRIT under more severe
conditions, i.e., we analyze it in terms of RMSE when at least two
of the four equal-powered Gaussian signals are strongly correlated,
as shown in the following signal correlation matrix $\bm R_{ss}$
\eqref{signals_correlated_matrix}:
\begin{equation}
\bm R_{ss}=\sigma_{s}^{2}\begin{bmatrix}
\label{signals_correlated_matrix}
1   &  0.9   & 0.6  & 0   \\
0.9 &   1    & 0.4  & 0.5    \\
0.6 &  0.4   & 1    & 0     \\
0   &  0.5   & 0    & 1
\end{bmatrix}.
\end{equation}

The signal-to-noise ratio $\text(SNR)$ is defined as $SNR\triangleq 10\log_{10}\left(\frac{\sigma_{s}^{2}}{\sigma_{n}^{2}} \right)$.
\begin{figure}[!h]
    \centering 
    \includegraphics[width=8cm, height=6cm]{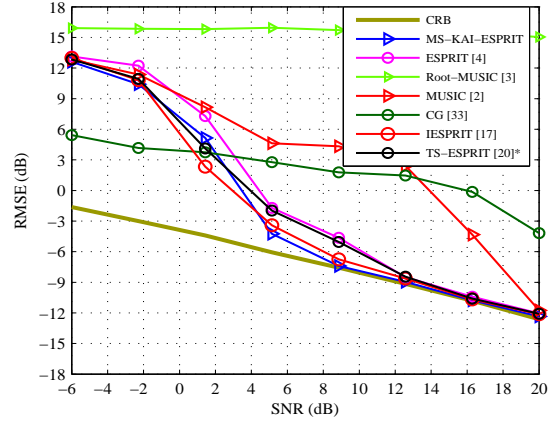} 
    \vspace{-1.0em}\caption{RMSE and the square root of CRB versus SNR with $P=4$ correlated sources, $M=40$, $N=25$, $L=250$ runs}
    \label{figura:RMSE_dB_correlated_2ponto4deg_40sens_25snap_250runs}
\end{figure}
In Fig.
\ref{figura:RMSE_dB_correlated_2ponto4deg_40sens_25snap_250runs}, we
can see the performance of the same algorithms plotted in Fig.
\ref{figura:DSP_RMSE_CRB_2ponto4deg_40sens_25snap_100runs} in terms
of $\mathrm{RMSE(dB)}$ versus SNR computed after $\mathrm{250}$
runs, when  the signal correlation matrix is given by
\eqref{signals_correlated_matrix}. As can be seen, the superior
performance of MS-KAI-ESPRIT occurs in the whole range between $4.0$
and $ 12$ dB , which can be considered a small but consistent gain.
From $ 12$dB to $ 20$dB MS-KAI-ESPRIT, TS-ESPRIT, IESPRIT and ESPRIT
have similar performance.  The values for which MS-KAI-ESPRIT is
outperformed are in the range between $ -6.0 $ and $4.0$dB.

\begin{figure}[!h]
    \centering 
    \includegraphics[width=8cm, height=6cm]{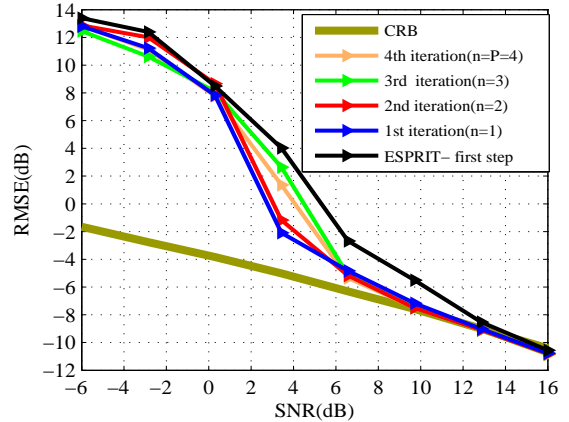} 
    \vspace{-1.0em}\caption{RMSE for each iteration of MS-KAI ESPRIT,original ESPRIT and CRB versus SNR with $P=4$ correlated sources, $M=40$, $N=25$, $L=200$ runs}
    \label{RMSE_dB_corr_letter_steps_full_reviewer1}
\end{figure}

\begin{figure}[!h]
    \centering 
    \includegraphics[width=8cm, height=6cm]{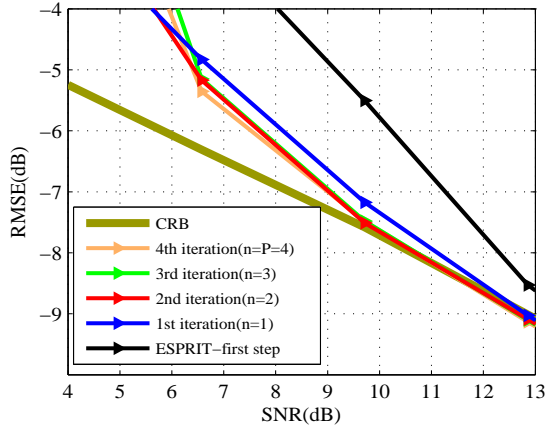} 
    \vspace{-1.0em}\caption{RMSE for each iteration of MS-KAI ESPRIT,original ESPRIT and CRB versus SNR with $P=4$ correlated sources, $M=40$, $N=25$, $L=200$ runs -magnification }
    \label{RMSE_dB_corr_letter_steps_reviewer1}
\end{figure}

In  Fig. \ref{RMSE_dB_corr_letter_steps_full_reviewer1}, we have
provided further  simulations to illustrate the performance of each
iteration of MS-KAI ESPRIT in terms of RMSE. The resulting
iterations can be compared to each other and to the the original
ESPRIT, which corresponds to the first step of MS-KAI ESPRIT. For
this purpose, we have considered the same scenario employed before,
except for the number of the trials, which is $L=200$ runs for  all
simulations. In particular, we have considered the case of
correlated sources. From Fig.
\ref{RMSE_dB_corr_letter_steps_reviewer1}, which is a magnified
detail of Fig. \ref{RMSE_dB_corr_letter_steps_full_reviewer1}, it
can be seen that the estimates become more accurate with the
increase of iterations.

%
%

\section{Conclusions}

We have proposed the MS-KAI-ESPRIT algorithm which exploits the
knowledge of source signals obtained on line and the structure of
the covariance matrix and its perturbations. An analytical study of
the MSE of this matrix free of side effects has shown that it is
less or equal than the MSE of the original matrix, resulting in
better performance of MS-KAI-ESPRIT especially in scenarios where
limited number of samples are available. The proposed MS-KAI-ESPRIT
algorithm can obtain significant gains in RMSE or probability of
resolution performance over previously reported techniques, and has
excellent potential for applications with short data records in
large-scale antenna systems for wireless communications, radar and
other large sensor arrays. The relatively high computational burden
required, which is associated with extra matrix multiplications, the
increment $\iota$ applied to reduce the undesirable side effects and
the iterations needed to progressively incorporate the knowledge
obtained on line as newer estimates can be justified for the
superior performance achieved. Future work will consider approaches
to reducing the computational cost.


\section*{{Appendix}}

{Here, we prove the inequality \eqref{problem_formulation2}
described in Section \ref{MSE_analysis}. We start by expressing the
MSE of the original data covariance matrix \eqref{covsample} as:}
\begin{align}
\mathrm{MSE}\left( \mathbf{\hat{R}}\right) =\mathbb E\left[\|\mathbf{\hat{R}}-
\mathbf{R}\|^2_F\right].
\label{problem_formulation3}
\end{align}
where $\mathbf{R} $ is the true covariance matrix .
Similarly, the  MSE of the data covariance matrix free of side effects $\mathbf{\hat{R}}^{(n+1)}$ can be expressed for the first iteration $n=1$ by making  use of  \eqref{modified_data_covariance}, as follows
\begin{align}
&\mathrm{MSE}\left( \mathbf{\hat{R}}^{(n+1)}\right)\big\rvert_{n=1}\nonumber\\&=\mathrm{MSE}\left( \mathbf{\hat{R}}^{(2)}\right)=\mathbb E\left[\|\mathbf{\hat{R}}^{(2)}-
\mathbf{R}\|^2_F\right]\nonumber\\
&=\mathbb E\left[\| \mathbf{\hat{R}}\:-\:\mathrm{\mu}\:(\mathbf{V}^{(1)}\:+\:\mathbf{V}^{(1)H})-
\mathbf{R}\|^2_F\right]\nonumber\\
&=\mathbb E\left[\|\left(  \mathbf{\hat{R}}-
\mathbf{R}\right) \:-\:\mathrm{\mu}\:(\mathbf{V}^{(1)}\:+\:\mathbf{V}^{(1)H})\|^2_F\right]
\label{MSE_eq1}
\end{align}
where for the sake of simplicity, from now on we omit the superscript $^{(1)}$, which refers to the first iteration.
In order to expand the result in \eqref{MSE_eq1}, we make use of the following proposition:

\medskip
\underline{Lemma 1}:
The squared Frobenius norm of the difference between any two matrices $\mathbf{A}$ $\in\mathbb{C}^{\mathit{m \times m}}$ and $\mathbf{B}$ $\in\mathbb{C}^{\mathit{m \times m}}$ is given by
\vspace{-0.5em}
\begin{align}
\|\mathbf{A}-
\mathbf{B}\|^2_F=\|\mathbf{A}\|^2_F +\|\mathbf{B}\|^2_F -\left( \Tr\mathbf{A}^{H}\mathbf{B} + \Tr\mathbf{A}\mathbf{B}^{H}\right)
\end{align}
\underline{Proof  of Lemma 1}:\\
The Frobenius norm of any $\mathbf{D}$ $\in$  $\mathbb{C}^{\mathit{m \times
        m}}$  matrix  is defined \cite{Vantrees} as \vspace{-1.00 em}
\begin{align}{
    \|\mathbf{D}\|_F=
    \left( \sum\limits_{i=1}^{m}\sum\limits_{j=1}^{m}\left| d_{ij}\right|^{2}\right )^\frac{1}{2}=\left[\Tr\left(\mathbf{ D}^{H}\mathbf{ D}\right)  \right]^{\frac{1}{2}}}
\label{Proof_Lemma_1}
\end{align}
We express $\mathbf{D} $ as a difference between  two  matrices $\mathbf{A}$ and  $\mathbf{B}$, both also $\in\mathbb{C}^{\mathit{m \times m}}$. Making use of Lemma1 and the properties of the trace, we obtain
{\begin{align}
    &\|\mathbf{A}-
    \mathbf{B}\|^2_F=  \Tr\left[\left( \mathbf{A}-\mathbf{B}\right)^{H} \left( \mathbf{A}-\mathbf{B}\right)\right]\nonumber\\
    &= \Tr\left[ \left( \mathbf{A}^{H}-\mathbf{B}^{H}\right) \left( \mathbf{A}-\mathbf{B}\right)\right]\nonumber\\
    &= \Tr\left[\left( \mathbf{A}^{H}\mathbf{A}\right)
    -\Tr\left( \mathbf{A}^{H}\mathbf{B}\right)- \Tr\left( \mathbf{B}^{H}\mathbf{A}\right)+\Tr\left( \mathbf{B}^{H}\mathbf{B}\right)\right]\nonumber\\
    &=\|\mathbf{A}\|^2_F +\|\mathbf{B}\|^2_F -\left( \Tr\mathbf{A}^{H}\mathbf{B} + \Tr\mathbf{A}\mathbf{B}^{H}\right),
    \end{align}}
which is the desired result.

Now, assuming that the
true $\mathbf{R}$ \cite{Haykin} and the  data covariance matrices  $\mathbf{\hat{R}} $ \cite{Haykin}  are Hermitian and  using \eqref{MSE_eq1}  combined with Lemma1, the cyclic \cite{Graybill} property of the trace  and the
linearity \cite{Karr} property of the expected value,  we get
\begin{align}
\mathrm{MSE}\left( \mathbf{\hat{R}}^{(2)}\right)& =\mathbb
E\left\lbrace \|\mathbf{\hat{R}}-
\mathbf{R}\|^2_F + \mathrm{\mu}^{2}\:\|\mathbf{V}\:+\:\mathbf{V}^{H}\|^2_F\right.\nonumber\\
&\left.-\Tr\left[ \left(\mathbf{\hat{R}}-
\mathbf{R}\right)^{H} \mu\left(\mathbf{V}\:+\:\mathbf{V}^{H} \right) \right]\right.\nonumber\\
& \left.-\Tr\left[\mu \left(\mathbf{V}+
\mathbf{V}^{H}\right)^{H} \left(\mathbf{\hat{R}}\:+\:\mathbf{R} \right)\right]\right\rbrace \nonumber\\
& = \mathbb E\left\lbrace \|\mathbf{\hat{R}}-
\mathbf{R}\|^2_F + \mathrm{\mu}^{2}\:\|\mathbf{V}\:+\:\mathbf{V}^{H}\|^2_F\right.\nonumber\\
&\left.-\mu\Tr\left[ \left(\mathbf{\hat{R}}-
\mathbf{R}\right)^{H} \left(\mathbf{V}\:+\:\mathbf{V}^{H} \right) \right]\right.\nonumber\\
& \left.-\mu\Tr\left[ \left(\mathbf{V}+
\mathbf{V}^{H}\right)^{H} \left(\mathbf{\hat{R}}\:+\:\mathbf{R} \right)\right]\right\rbrace \nonumber\\
&=\mathbb E\left\lbrace \|\mathbf{\hat{R}}-
\mathbf{R}\|^2_F + \mathrm{\mu}^{2}\:\|\mathbf{V}\:+\:\mathbf{V}^{H}\|^2_F\right.\nonumber\\
&\left.-\mu\Tr\left[ \left(\mathbf{\hat{R}}-
\mathbf{R}\right) \left(\mathbf{V}\:+\:\mathbf{V}^{H} \right) \right]\right.\nonumber\\
& \left.-\mu\Tr\left[  \left(\mathbf{V}^{H}+
\mathbf{V}\right)\left(\mathbf{\hat{R}}\:+\:\mathbf{R} \right)\right]\right\rbrace \nonumber\\
&=\mathbb E\left\lbrace \|\mathbf{\hat{R}}-
\mathbf{R}\|^2_F + \mathrm{\mu}^{2}\:\|\mathbf{V}\:+\:\mathbf{V}^{H}\|^2_F\right.\nonumber\\
&\left.-\mu\Tr\left[ \left(\mathbf{\hat{R}}-
\mathbf{R}\right) \left(\mathbf{V}\:+\:\mathbf{V}^{H} \right) \right]\right.\nonumber\\
& \left.-\mu\Tr\left[  \left(\mathbf{\hat{R}}\:+\:\mathbf{R} \right)\left(\mathbf{V}+
\mathbf{V}^{H}\right)\right]\right\rbrace\nonumber\\
&=\mathbb E\left\lbrace \|\mathbf{\hat{R}}-
\mathbf{R}\|^2_F\right\rbrace  +\mathrm{\mu}^{2} \mathbb E\left\lbrace\|\mathbf{V}\:+\:\mathbf{V}^{H}\|^2_F\right\rbrace \mathbb \nonumber\\
&-2\mu\mathbb E\left\lbrace \Tr\left[ \left(\mathbf{\hat{R}}-
\mathbf{R}\right) \left(\mathbf{V}\:+\:\mathbf{V}^{H} \right) \right]\right\rbrace\nonumber\\
&=\mathrm{MSE}\left( \mathbf{\hat{R}}\right) +\mathrm{\mu}^{2} \mathbb E\left\lbrace\|\mathbf{V}\:+\:\mathbf{V}^{H}\|^2_F\right\rbrace \mathbb \nonumber\\
&-2\mu\mathbb E\left\lbrace \Tr\left[ \left(\mathbf{\hat{R}}-
\mathbf{R}\right) \left(\mathbf{V}\:+\:\mathbf{V}^{H} \right) \right]\right\rbrace
\label{MSE_eq2}
\end{align}
By moving the first summand of \eqref{MSE_eq2} to its first element,
we obtain the intended expression for the difference between the
$MSEs$ of the data covariance matrix free of perturbations and the
original one, i.e.:

\begin{align}
\mathrm{MSE}\left( \mathbf{\hat{R}}^{(n+1)}\right)\big\rvert_{n=1}-\mathrm{MSE}\left( \mathbf{\hat{R}}\right)=\mathrm{\mu}^{2} \mathbb E\left\lbrace\|\mathbf{V}+\mathbf{V}^{H}\|^2_F\right\rbrace \mathbb \nonumber\\
\quad -2\mu\mathbb E\left\lbrace \Tr\left[ \left(\mathbf{\hat{R}}-
\mathbf{R}\right) \left(\mathbf{V}\:+\:\mathbf{V}^{H} \right)
\right]\right\rbrace.
\label{Difference_MSEs}
\end{align}

Now, we  expand the expressions inside braces of the second member of \eqref{Difference_MSEs} individually. We start with the first summand

\begin{align}
\|\mathbf{V}+
\mathbf{V}^{H}\|^2_F& =\|\mathbf{V}\|^2_F +\|\mathbf{V}^{H}\|^2_F +\Tr\left(\mathbf{V}^{H}\mathbf{V}^{H} \right)+\nonumber\\
& \quad \Tr\left(\mathbf{(V^{\mathit{H}})}^{H}\mathbf{V} \right)\nonumber\\
&=\|\mathbf{V}\|^2_F +\|\mathbf{V}^{H}\|^2_F
+\Tr\left(\mathbf{V}^{H}\mathbf{V}^{H} \right)+
\Tr\left(\mathbf{V}\mathbf{V} \right).
\label{By_products}
\end{align}

The equation \eqref{By_products} can be computed by using the
projection matrices of the signal and the noise subspaces and the
data covariance matrix by using \eqref{terms_deducted},
\eqref{noise_projection}, the idempotence \cite{Vantrees} \cite{Graybill} of $\mathbf{\hat{Q}}_{A}$
and the cyclic property \cite{Graybill} of the trace. Starting with the computation
of its fourth summand, we have

\begin{align}
\Tr\left(\mathbf{V}\mathbf{V} \right)& = \Tr\left[\left( \mathbf{\hat{Q}}_{A}\:\mathbf{\hat{R}}\:\mathbf{\hat{Q}}_{A}^{\perp} \right)\left( \mathbf{\hat{Q}}_{A}\:\mathbf{\hat{R}}\:\mathbf{\hat{Q}}_{A}^{\perp} \right)\right] \nonumber\\
&= \Tr\left[  \mathbf{\hat{Q}}_{A}\:\mathbf{\hat{R}}\left(\mathbf{I_{M}}-\mathbf{\hat{Q}}_{A} \right)\: \mathbf{\hat{Q}}_{A}\:\mathbf{\hat{R}}\left(\mathbf{I_{M}}-\mathbf{\hat{Q}}_{A} \right) \right]\nonumber\\
&=\Tr\left[\left( \mathbf{\hat{Q}}_{A}\:\mathbf{\hat{R}}- \mathbf{\hat{Q}}_{A}\:\mathbf{\hat{R}}\:\mathbf{\hat{Q}}_{A}\right)\right.\nonumber\\
&\left.\left( \mathbf{\hat{Q}}_{A}\:\mathbf{\hat{R}}- \mathbf{\hat{Q}}_{A}\:\mathbf{\hat{R}}\:\mathbf{\hat{Q}}_{A}\right) \right]\nonumber\\
&= \Tr\left[\mathbf{\hat{Q}}_{A}\:\mathbf{\hat{R}}\:\mathbf{\hat{Q}}_{A}\:\mathbf{\hat{R}}  -\mathbf{\hat{Q}}_{A}\:\mathbf{\hat{R}}\:\mathbf{\hat{Q}}_{A}\:\mathbf{\hat{R}}\:\mathbf{\hat{Q}}_{A}\right.\nonumber\\
&\left.-\mathbf{\hat{Q}}_{A}\:\mathbf{\hat{R}}\:\mathbf{\hat{Q}}_{A}\mathbf{\hat{Q}}_{A}\:\mathbf{\hat{R}} +\mathbf{\hat{Q}}_{A}\:\mathbf{\hat{R}}\:\mathbf{\hat{Q}}_{A}\mathbf{\hat{Q}}_{A}\:\mathbf{\hat{R}}\:\mathbf{\hat{Q}}_{A}\right] \nonumber\\
&=\Tr\left(\mathbf{\hat{Q}}_{A}\:\mathbf{\hat{R}}\:\mathbf{\hat{Q}}_{A}\:\mathbf{\hat{R}}\right)   -\Tr\left(\mathbf{\hat{Q}}_{A}\:\mathbf{\hat{R}}\:\mathbf{\hat{Q}}_{A}\:\mathbf{\hat{R}}\:\mathbf{\hat{Q}}_{A}\right) \nonumber\\
&-\Tr\left(\mathbf{\hat{Q}}_{A}\:\mathbf{\hat{R}}\:\mathbf{\hat{Q}}_{A}\mathbf{\hat{Q}}_{A}\:\mathbf{\hat{R}}\right)\nonumber\\  & +\Tr\left(\mathbf{\hat{Q}}_{A}\:\mathbf{\hat{R}}\:\mathbf{\hat{Q}}_{A}\mathbf{\hat{Q}}_{A}\:\mathbf{\hat{R}}\:\mathbf{\hat{Q}}_{A}\right)\nonumber\\
&= \Tr\left(\mathbf{\hat{Q}}_{A}\:\mathbf{\hat{R}}\:\mathbf{\hat{Q}}_{A}\:\mathbf{\hat{R}}\right)-\Tr\left(\mathbf{\hat{Q}}_{A}\:\mathbf{\hat{R}}\:\mathbf{\hat{Q}}_{A}\:\mathbf{\hat{R}}\right)\nonumber\\
&-\Tr\left(\mathbf{\hat{Q}}_{A}\:\mathbf{\hat{R}}\:\mathbf{\hat{Q}}_{A}\:\mathbf{\hat{R}}\right)+\Tr\left(\mathbf{\hat{Q}}_{A}\:\mathbf{\hat{R}}\:\mathbf{\hat{Q}}_{A}\:\mathbf{\hat{R}}\right)=0.
\label{Trace_VV}
\end{align}
Taking into account that the data covariance matrix $ \mathbf{\hat{R}} $ and the estimate of the projection matrix of the noise subspace  $ \mathbf{\hat{Q}}_{A}^{\perp} $ are Hermitian, we can evaluate the third summand of \eqref{By_products}  as follows:
\begin{align}
&\Tr\left(\mathbf{V}^{H}\mathbf{V}^{H} \right)= \Tr\left[\left( \mathbf{\hat{Q}}_{A}\:\mathbf{\hat{R}}\:\mathbf{\hat{Q}}_{A}^{\perp} \right)^{H}\left( \mathbf{\hat{Q}}_{A}\:\mathbf{\hat{R}}\:\mathbf{\hat{Q}}_{A}^{\perp} \right)^{H}\right] \nonumber\\
&= \Tr\left\lbrace \left[ \left( \mathbf{\hat{Q}}_{A}^{\perp}\right)^{H}  \:\mathbf{\hat{R}}^{H}\:\mathbf{\hat{Q}}_{A}^{H} \right] \left[ \left( \mathbf{\hat{Q}}_{A}^{\perp}\right)^{H}  \:\mathbf{\hat{R}}^{H}\:\mathbf{\hat{Q}}_{A}^{H} \right] \right\rbrace \nonumber\\
&= \Tr\left\lbrace \left[ \mathbf{\hat{Q}}_{A}^{\perp} \:\mathbf{\hat{R}}\:\mathbf{\hat{Q}}_{A} \right] \left[ \mathbf{\hat{Q}}_{A}^{\perp} \:\mathbf{\hat{R}}\:\mathbf{\hat{Q}}_{A} \right] \right\rbrace  \nonumber\\
&= \Tr\left\lbrace \left[  \mathbf{\left(I_{M}-\mathbf{\hat{Q}}_{A} \right)\mathbf{\hat{R}}\:\hat{Q}}_{A}\right] \left[  \:\mathbf{\left(I_{M}-\mathbf{\hat{Q}}_{A} \right)\mathbf{\hat{R}}\:\hat{Q}}_{A} \right]\right\rbrace \nonumber\\
&=\Tr\left\lbrace \left[ \mathbf{\hat{R}}\:\mathbf{\hat{Q}}_{A}- \mathbf{\hat{Q}}_{A}\:\mathbf{\hat{R}}\:\mathbf{\hat{Q}}_{A}\right]\left[ \mathbf{\hat{R}}\:\mathbf{\hat{Q}}_{A}- \mathbf{\hat{Q}}_{A}\:\mathbf{\hat{R}}\:\mathbf{\hat{Q}}_{A}\right] \right\rbrace
\nonumber\\
&=\Tr\left\lbrace \mathbf{\hat{R}}\:\mathbf{\hat{Q}}_{A}\:\mathbf{\hat{R}}\:\mathbf{\hat{Q}}_{A}  -\mathbf{\hat{R}}\:\mathbf{\hat{Q}}_{A}\mathbf{\hat{Q}}_{A}\:\:\mathbf{\hat{R}}\:\mathbf{\hat{Q}}_{A}\right.\nonumber\\
&\left.-\mathbf{\hat{Q}}_{A}\:\mathbf{\hat{R}}\:\mathbf{\hat{Q}}_{A}\:\mathbf{\hat{R}}\:\mathbf{\hat{Q}}_{A} +\mathbf{\hat{Q}}_{A}\:\mathbf{\hat{R}}\:\mathbf{\hat{Q}}_{A}\mathbf{\hat{Q}}_{A}\:\mathbf{\hat{R}}\:\mathbf{\hat{Q}}_{A}\right\rbrace  \nonumber\\
&=\Tr\left(\mathbf{\hat{R}}\:\mathbf{\hat{Q}}_{A}\:\mathbf{\hat{R}}\:\mathbf{\hat{Q}}_{A}\:\right)   -\Tr\left(\mathbf{\hat{R}}\:\mathbf{\hat{Q}}_{A}\mathbf{\hat{Q}}_{A}\:\mathbf{\hat{R}}\:\mathbf{\hat{Q}}_{A}\right) \nonumber\\
&-\Tr\left(\mathbf{\hat{Q}}_{A}\:\mathbf{\hat{R}}\:\mathbf{\hat{Q}}_{A}\:\mathbf{\hat{R}}\:\mathbf{\hat{Q}}_{A}\right)  +\Tr\left(\mathbf{\hat{Q}}_{A}\:\mathbf{\hat{R}}\:\mathbf{\hat{Q}}_{A}\mathbf{\hat{Q}}_{A}\:\mathbf{\hat{R}}\:\mathbf{\hat{Q}}_{A}\right)\nonumber\\
&= \Tr\left(\mathbf{\hat{R}}\:\mathbf{\hat{Q}}_{A}\:\mathbf{\hat{R}}\:\mathbf{\hat{Q}}_{A}\right)-\Tr\left(\mathbf{\hat{R}}\:\mathbf{\hat{Q}}_{A}\:\mathbf{\hat{R}}\mathbf{\hat{Q}}_{A}\right)\nonumber\\
&-\Tr\left(\mathbf{\hat{Q}}_{A}\:\mathbf{\hat{R}}\:\mathbf{\hat{Q}}_{A}\:\mathbf{\hat{R}}\right)+\Tr\left(\mathbf{\hat{Q}}_{A}\:\mathbf{\hat{R}}\:\mathbf{\hat{Q}}_{A}\:\mathbf{\hat{R}}\right)=0.
\label{Trace_VHVH}
\end{align}
By using \eqref{Proof_Lemma_1}, we can expand the first and the second summands  of
\eqref{By_products} as follows:
\begin{align}
&\|\mathbf{V}\|^2_F +\|\mathbf{V}^{H}\|^2_F =\Tr\left(\mathbf{V}^{H}\mathbf{V} \right)+\Tr\left(\left( \mathbf{V}^{H}\right)^{H} \mathbf{V}^{H} \right)\nonumber\\
&= \Tr\left(\mathbf{V}^{H}\mathbf{V} \right)+\Tr\left(\mathbf{V}\mathbf{V}^{H} \right)\nonumber\\
&=\Tr\left(\mathbf{V}\mathbf{V}^{H}
\right)+\Tr\left(\mathbf{V}\mathbf{V}^{H}
\right)=2\Tr\left(\mathbf{V}\mathbf{V}^{H} \right).
\label{Sq_Frob_norm_of_VmaisVH}
\end{align}
Equation \eqref{Sq_Frob_norm_of_VmaisVH} can be  expressed in terms
of the projection matrices of the signal and the noise subspaces and
the data covariance, in a similar way as for the third and fourth
summands of \eqref{By_products}, as follows:
\begin{align}
& 2\Tr\left(\mathbf{V}\mathbf{V}^{H} \right)=2\Tr\left[ \left( \mathbf{\hat{Q}}_{A}\:\mathbf{\hat{R}}\:\mathbf{\hat{Q}}_{A}^{\perp} \right) \left( \mathbf{\hat{Q}}_{A}\:\mathbf{\hat{R}}\:\mathbf{\hat{Q}}_{A}^{\perp} \right)^{H}\right]\nonumber\\
&=2\Tr\left\lbrace\mathbf{\hat{Q}}_{A}\mathbf{\hat{R}} \left(\mathbf{ I_{M} }-\mathbf{\hat{Q}}_{A}  \right) \left[ \mathbf{\hat{Q}}_{A}\mathbf{\hat{R}} \left(\mathbf{ I_{M} }-\mathbf{\hat{Q}}_{A} \right)\right]^{H} \right\rbrace \nonumber\\
&=2\Tr\left\lbrace\left(\mathbf{\hat{Q}}_{A}\mathbf{\hat{R}}- \mathbf{\hat{Q}}_{A}\mathbf{\hat{R}}\mathbf{\hat{Q}}_{A}\right)\left(\mathbf{\hat{Q}}_{A}\mathbf{\hat{R}}- \mathbf{\hat{Q}}_{A}\mathbf{\hat{R}}\mathbf{\hat{Q}}_{A}\right)^{H}   \right\rbrace \nonumber\\
&=2\Tr\left\lbrace\mathbf{\hat{Q}}_{A}\mathbf{\hat{R}}\mathbf{\hat{R}}\mathbf{\hat{Q}}_{A}- \mathbf{\hat{Q}}_{A}\mathbf{\hat{R}}\mathbf{\hat{Q}}_{A}\mathbf{\hat{R}}\right.\nonumber\\
&\left.- \mathbf{\hat{Q}}_{A}\mathbf{\hat{R}}\mathbf{\hat{Q}}_{A}\mathbf{\hat{R}}\mathbf{\hat{Q}}_{A}+ \mathbf{\hat{Q}}_{A}\mathbf{\hat{R}}\mathbf{\hat{Q}}_{A}\mathbf{\hat{Q}}_{A}\mathbf{\hat{R}} \right\rbrace \nonumber\\
&=2\left\lbrace\Tr\left(\mathbf{\hat{Q}}_{A}\mathbf{\hat{R}}\mathbf{\hat{R}}\mathbf{\hat{Q}}_{A} \right) -\Tr\left(\mathbf{\hat{Q}}_{A}\mathbf{\hat{R}}\mathbf{\hat{Q}}_{A}\mathbf{\hat{R}} \right)  \right.\nonumber\\
&\left.-\Tr\left( \mathbf{\hat{Q}}_{A}\mathbf{\hat{R}}\mathbf{\hat{Q}}_{A}\mathbf{\hat{R}}\mathbf{\hat{Q}}_{A}\right)  +\Tr\left(\mathbf{\hat{Q}}_{A}\mathbf{\hat{R}}\mathbf{\hat{Q}}_{A}\mathbf{\hat{Q}}_{A}\mathbf{\hat{R}} \right)   \right\rbrace \nonumber\\
&=2\left\lbrace\Tr\left(\mathbf{\hat{Q}}_{A}\mathbf{\hat{Q}}_{A}\mathbf{\hat{R}}\mathbf{\hat{R}} \right) -\Tr\left(\mathbf{\hat{Q}}_{A}\mathbf{\hat{R}}\mathbf{\hat{Q}}_{A}\mathbf{\hat{R}} \right)  \right.\nonumber\\
&\left.-\Tr\left( \mathbf{\hat{Q}}_{A}\mathbf{\hat{R}}\mathbf{\hat{Q}}_{A}\mathbf{\hat{R}}\right)  +\Tr\left(\mathbf{\hat{Q}}_{A}\mathbf{\hat{R}}\mathbf{\hat{Q}}_{A}\mathbf{\hat{R}} \right)   \right\rbrace \nonumber\\
&=2\left\lbrace\Tr\left(\mathbf{\hat{Q}}_{A}\mathbf{\hat{Q}}_{A}\mathbf{\hat{R}}\mathbf{\hat{R}} \right) -\Tr\left(\mathbf{\hat{Q}}_{A}\mathbf{\hat{R}}\mathbf{\hat{Q}}_{A}\mathbf{\hat{R}} \right)\right\rbrace
\label{Double_tr_V_mais_VH}
\end{align}
From \eqref{By_products}, \eqref{Trace_VV}, \eqref{Trace_VHVH}, \eqref{Sq_Frob_norm_of_VmaisVH} and \eqref{Double_tr_V_mais_VH},  we obtain the first summand of \eqref{Difference_MSEs}, as follows:
\begin{align}
\mathrm{\mu}^{2} \mathbb {E}\left\lbrace\|\mathbf{V}+\mathbf{V}^{H}\|^2_F\right\rbrace& =\mathrm{2\mu}^{2} \mathbb{E}\left\lbrace\Tr\left(\mathbf{\hat{Q}}_{A}\mathbf{\hat{Q}}_{A}\mathbf{\hat{R}}\mathbf{\hat{R}} \right) \right.\nonumber\\
&\left.-\Tr\left(\mathbf{\hat{Q}}_{A}\mathbf{\hat{R}}\mathbf{\hat{Q}}_{A}\mathbf{\hat{R}} \right)\right\rbrace
\label{First_summ_Difference_MSEs_key}
\end{align}
In order to finish the expansion of the  expressions inside braces of the second member of \eqref{Difference_MSEs}, now  we deal with its second summand, in which we make use of the cyclic property \cite{Graybill} of the trace and the idempotence \cite{Vantrees} \cite{Graybill} of $\mathbf{\hat{Q}}_{A}$.

\begin{align}
& \Tr\left[\left( \mathbf{\hat{R}}-\mathbf{R}\right)  \left( \mathbf{V}+\mathbf{V}^{H}\right) \right] =\left\lbrace \Tr\left( \mathbf{\hat{R}}-\mathbf{R}\right)\right.\nonumber\\
&\left.\left[  \mathbf{\hat{Q}}_{A}\:\mathbf{\hat{R}}\:\mathbf{\hat{Q}}_{A}^{\perp}  + \left( \mathbf{\hat{Q}}_{A}\:\mathbf{\hat{R}}\:\mathbf{\hat{Q}}_{A}^{\perp} \right)^{H}\right]\right\rbrace \nonumber\\
&=\Tr\left\lbrace\left( \mathbf{\hat{R}}-\mathbf{R}\right)\left[ \mathbf{\hat{Q}}_{A}\mathbf{\hat{R}} \left(\mathbf{ I_{M} }-\mathbf{\hat{Q}}_{A} \right)\right.\right.\nonumber\\
&\left.\left.+ \left(  \mathbf{\hat{Q}}_{A}\mathbf{\hat{R}} \left(\mathbf{ I_{M} }-\mathbf{\hat{Q}}_{A}  \right)\right)^{H} \right] \right\rbrace \nonumber\\
&=\Tr\left\lbrace\left( \mathbf{\hat{R}}-\mathbf{R}\right)\left[\mathbf{\hat{Q}}_{A}\mathbf{\hat{R}}-\mathbf{\hat{Q}}_{A}\mathbf{\hat{R}}\mathbf{\hat{Q}}_{A}\right.\right.\nonumber\\
&\left.\left.+\left( \mathbf{\hat{Q}}_{A}\mathbf{\hat{R}}-\mathbf{\hat{Q}}_{A}\mathbf{\hat{R}}\mathbf{\hat{Q}}_{A}\right)^{H}  \right]  \right\rbrace \nonumber\\
&=\Tr\left\lbrace\left( \mathbf{\hat{R}}-\mathbf{R}\right)\left[\mathbf{\hat{Q}}_{A}\mathbf{\hat{R}}-\mathbf{\hat{Q}}_{A}\mathbf{\hat{R}}\mathbf{\hat{Q}}_{A}+\mathbf{\hat{R}\mathbf{\hat{Q}}_{A}}-\mathbf{\hat{Q}}_{A}\mathbf{\hat{R}}\mathbf{\hat{Q}}_{A} \right]  \right\rbrace \nonumber\\
&=\Tr\left\lbrace \mathbf{\hat{R}}\mathbf{\hat{Q}}_{A}\mathbf{\hat{R}}+\mathbf{\hat{R}}\mathbf{\hat{R}\mathbf{\hat{Q}}_{A}}-2\mathbf{\hat{R}}\mathbf{\hat{Q}}_{A}\mathbf{\hat{R}}\mathbf{\hat{Q}}_{A} \right.\nonumber\\
&\left.-\mathbf{R}\mathbf{\hat{Q}}_{A}\mathbf{\hat{R}}-\mathbf{R}\mathbf{\hat{R}\mathbf{\hat{Q}}_{A}}+2\mathbf{R}\mathbf{\hat{Q}}_{A}\mathbf{\hat{R}}\mathbf{\hat{Q}}_{A}\right\rbrace \nonumber\\
&=\Tr \mathbf{\hat{R}}\mathbf{\hat{Q}}_{A}\mathbf{\hat{R}}+\Tr\mathbf{\hat{R}}\mathbf{\hat{R}\mathbf{\hat{Q}}_{A}}-2\Tr\mathbf{\hat{R}}\mathbf{\hat{Q}}_{A}\mathbf{\hat{R}}\mathbf{\hat{Q}}_{A} \nonumber\\
&-\Tr\mathbf{R}\mathbf{\hat{Q}}_{A}\mathbf{\hat{R}}-\Tr\mathbf{R}\mathbf{\hat{R}\mathbf{\hat{Q}}_{A}}+2\Tr\mathbf{R}\mathbf{\hat{Q}}_{A}\mathbf{\hat{R}}\mathbf{\hat{Q}}_{A}\nonumber\\
&=\Tr\mathbf{\hat{Q}}_{A}\mathbf{\hat{R}}\mathbf{\hat{R}}+\Tr\mathbf{\hat{Q}}_{A}\mathbf{\hat{R}}\mathbf{\hat{R}}-2\Tr\mathbf{\hat{Q}}_{A}\mathbf{\hat{R}}\mathbf{\hat{Q}}_{A}\mathbf{\hat{R}} \nonumber\\
&-\Tr\mathbf{R}\mathbf{\hat{Q}}_{A}\mathbf{\hat{R}}-\Tr\mathbf{\hat{Q}}_{A}\mathbf{R}\mathbf{\hat{R}}+2\Tr\mathbf{\hat{Q}}_{A}\mathbf{R}\mathbf{\hat{Q}}_{A}\mathbf{\hat{R}}\nonumber\\
&=2\Tr\mathbf{\hat{Q}}_{A}\mathbf{\hat{R}}\mathbf{\hat{R}}-2\Tr\mathbf{\hat{Q}}_{A}\mathbf{\hat{R}}\mathbf{\hat{Q}}_{A}\mathbf{\hat{R}}-\Tr\mathbf{R}\mathbf{\hat{Q}}_{A}\mathbf{\hat{R}} \nonumber\\
&-\Tr\mathbf{\hat{Q}}_{A}\mathbf{R}\mathbf{\hat{R}}+2\Tr\mathbf{\hat{Q}}_{A}\mathbf{R}\mathbf{\hat{Q}}_{A}\mathbf{\hat{R}}\nonumber\\
&=2\Tr\mathbf{\hat{Q}}_{A}\mathbf{\hat{Q}}_{A}\mathbf{\hat{R}}\mathbf{\hat{R}}-2\Tr\mathbf{\hat{Q}}_{A}\mathbf{\hat{R}}\mathbf{\hat{Q}}_{A}\mathbf{\hat{R}}-\Tr\mathbf{R}\mathbf{\hat{Q}}_{A}\mathbf{\hat{Q}}_{A}\mathbf{\hat{R}} \nonumber\\
&-\Tr\mathbf{\hat{Q}}_{A}\mathbf{\hat{Q}}_{A}\mathbf{R}\mathbf{\hat{R}}+2\Tr\mathbf{\hat{Q}}_{A}\mathbf{R}\mathbf{\hat{Q}}_{A}\mathbf{\hat{R}}
\label{tr_Rsample_R_V_mais_VH}
\end{align}
By using \eqref{tr_Rsample_R_V_mais_VH}, we can straightforwardly write
the second summand of the second member of \eqref{Difference_MSEs} in terms of the projection matrices of the signal and the noise subspaces and the data covariance matrix as follows:
\begin{align}
&-2\mu\mathbb E\left\lbrace \Tr\left[ \left(\mathbf{\hat{R}}-
\mathbf{R}\right) \left(\mathbf{V}\:+\:\mathbf{V}^{H} \right) \right]\right\rbrace
\nonumber\\
&=-2\mu\mathbb E\left\lbrace 2\Tr\mathbf{\hat{Q}}_{A}\mathbf{\hat{Q}}_{A}\mathbf{\hat{R}}\mathbf{\hat{R}}-2\Tr\mathbf{\hat{Q}}_{A}\mathbf{\hat{R}}\mathbf{\hat{Q}}_{A}\mathbf{\hat{R}}-\Tr\mathbf{R}\mathbf{\hat{Q}}_{A}\mathbf{\hat{Q}}_{A}\mathbf{\hat{R}}\right. \nonumber\\
&\left.-\Tr\mathbf{\hat{Q}}_{A}\mathbf{\hat{Q}}_{A}\mathbf{R}\mathbf{\hat{R}}+2\Tr\mathbf{\hat{Q}}_{A}\mathbf{R}\mathbf{\hat{Q}}_{A}\mathbf{\hat{R}} \right\rbrace\nonumber\\
&= -4\mu\mathbb E\left\lbrace \Tr\mathbf{\hat{Q}}_{A}\mathbf{\hat{Q}}_{A}\mathbf{\hat{R}}\mathbf{\hat{R}}-\Tr\mathbf{\hat{Q}}_{A}\mathbf{\hat{R}}\mathbf{\hat{Q}}_{A}\mathbf{\hat{R}}\right\rbrace\nonumber\\
& -2\mu\left\lbrace-\Tr\mathbb E\left[ \mathbf{R}\mathbf{\hat{Q}}_{A}\mathbf{\hat{Q}}_{A}\mathbf{\hat{R}}\right]-\Tr\mathbb E\left[ \mathbf{\hat{Q}}_{A}\mathbf{\hat{Q}}_{A}\mathbf{R}\mathbf{\hat{R}}  \right]\right. \nonumber\\
&\left.+2\Tr\mathbb E\left[ \mathbf{\hat{Q}}_{A}\mathbf{R}\mathbf{\hat{Q}}_{A}\mathbf{\hat{R}}  \right]    \right\rbrace \nonumber\\
&= -4\mu\mathbb E\left\lbrace \Tr\mathbf{\hat{Q}}_{A}\mathbf{\hat{Q}}_{A}\mathbf{\hat{R}}\mathbf{\hat{R}}-\Tr\mathbf{\hat{Q}}_{A}\mathbf{\hat{R}}\mathbf{\hat{Q}}_{A}\mathbf{\hat{R}}\right\rbrace\nonumber\\
& -2\mu\left\lbrace-\Tr\mathbb \mathbf{R}\mathbf{\hat{Q}}_{A}\mathbf{\hat{Q}}_{A} \mathbb E\left[\mathbf{\hat{R}}\right]-\Tr \mathbf{\hat{Q}}_{A}\mathbf{\hat{Q}}_{A}\mathbf{R}\mathbb E\left[\mathbf{\hat{R}}  \right]\right. \nonumber\\
&\left.+2\Tr\mathbf{\hat{Q}}_{A}\mathbf{R}\mathbf{\hat{Q}}_{A} \mathbb E\left[\mathbf{\hat{R}}\right]    \right\rbrace
\label{Sec_summ_Difference_MSEs_key}
\end{align}
Now, by using \eqref{First_summ_Difference_MSEs_key} and
\eqref{Sec_summ_Difference_MSEs_key}, and assuming that $\mathbb
E\left[\mathbf{\hat{R}}\right] $ is an unbiased estimate of
$\mathbf{\hat{R}}$, i.e., $\mathbb E\left[\mathbf{\hat{R}}\right]
=\mathbf{R}$, we can rewrite \eqref{Difference_MSEs} as follows:
\begin{align}
&\mathrm{MSE}\left( \mathbf{\hat{R}}^{(n+1)}\right)\big\rvert_{n=1}-\mathrm{MSE}\left( \mathbf{\hat{R}}\right)=\mathrm{\mu}^{2} \mathbb E\left\lbrace\|\mathbf{V}+\mathbf{V}^{H}\|^2_F\right\rbrace \mathbb \nonumber\\
&-2\mu\mathbb E\left\lbrace \Tr\left[ \left(\mathbf{\hat{R}}-
\mathbf{R}\right) \left(\mathbf{V}\:+\:\mathbf{V}^{H} \right) \right]\right\rbrace \nonumber\\
&=\mathrm{2\mu}^{2} \mathbb{E}\left\lbrace\Tr\mathbf{\hat{Q}}_{A}\mathbf{\hat{Q}}_{A}\mathbf{\hat{R}}\mathbf{\hat{R}}  -\Tr\mathbf{\hat{Q}}_{A}\mathbf{\hat{R}}\mathbf{\hat{Q}}_{A}\mathbf{\hat{R}} \right\rbrace\nonumber\\
&-4\mu\mathbb E\left\lbrace \Tr\mathbf{\hat{Q}}_{A}\mathbf{\hat{Q}}_{A}\mathbf{\hat{R}}\mathbf{\hat{R}}-\Tr\mathbf{\hat{Q}}_{A}\mathbf{\hat{R}}\mathbf{\hat{Q}}_{A}\mathbf{\hat{R}}\right\rbrace\nonumber\\
& -2\mu\left\lbrace-\Tr \mathbf{R}\mathbf{\hat{Q}}_{A}\mathbf{\hat{Q}}_{A} \mathbf{R}-\Tr\mathbf{\hat{Q}}_{A}\mathbf{\hat{Q}}_{A}\mathbf{R}\mathbf{R}\right. \nonumber\\
&\left.+2\Tr\mathbf{\hat{Q}}_{A}\mathbf{R}\mathbf{\hat{Q}}_{A} \mathbf{R}    \right\rbrace\nonumber\\
&=\mathrm{2\mu}^{2} \mathbb{E}\left\lbrace\Tr\mathbf{\hat{Q}}_{A}\mathbf{\hat{Q}}_{A}\mathbf{\hat{R}}\mathbf{\hat{R}}  -\Tr\mathbf{\hat{Q}}_{A}\mathbf{\hat{R}}\mathbf{\hat{Q}}_{A}\mathbf{\hat{R}} \right\rbrace\nonumber\\
&-4\mu\mathbb E\left\lbrace \Tr\mathbf{\hat{Q}}_{A}\mathbf{\hat{Q}}_{A}\mathbf{\hat{R}}\mathbf{\hat{R}}-\Tr\mathbf{\hat{Q}}_{A}\mathbf{\hat{R}}\mathbf{\hat{Q}}_{A}\mathbf{\hat{R}}\right\rbrace\nonumber\\
& -2\mu\left\lbrace-2\Tr \mathbf{R}\mathbf{\hat{Q}}_{A}\mathbf{\hat{Q}}_{A} \mathbf{R} +2\Tr\mathbf{\hat{Q}}_{A}\mathbf{R}\mathbf{\hat{Q}}_{A} \mathbf{R}    \right\rbrace\nonumber\\
&=\mathrm{2\mu}^{2} \mathbb{E}\left\lbrace\Tr\mathbf{\hat{Q}}_{A}\mathbf{\hat{Q}}_{A}\mathbf{\hat{R}}\mathbf{\hat{R}}  -\Tr\mathbf{\hat{Q}}_{A}\mathbf{\hat{R}}\mathbf{\hat{Q}}_{A}\mathbf{\hat{R}} \right\rbrace\nonumber\\
&-4\mu\mathbb E\left\lbrace \Tr\mathbf{\hat{Q}}_{A}\mathbf{\hat{Q}}_{A}\mathbf{\hat{R}}\mathbf{\hat{R}}-\Tr\mathbf{\hat{Q}}_{A}\mathbf{\hat{R}}\mathbf{\hat{Q}}_{A}\mathbf{\hat{R}}\right\rbrace\nonumber\\
& -4\mu\left\lbrace\Tr \mathbf{\hat{Q}}_{A}\mathbf{\hat{Q}}_{A} \mathbf{R}\mathbf{R} -\Tr\mathbf{\hat{Q}}_{A}\mathbf{R}\mathbf{\hat{Q}}_{A} \mathbf{R} \right\rbrace\nonumber\\
&=\left( \mathrm{2\mu}^{2}-\mathrm{4\mu}\right)  \mathbb{E}\left\lbrace\Tr\mathbf{\hat{Q}}_{A}\mathbf{\hat{Q}}_{A}\mathbf{\hat{R}}\mathbf{\hat{R}}  -\Tr\mathbf{\hat{Q}}_{A}\mathbf{\hat{R}}\mathbf{\hat{Q}}_{A}\mathbf{\hat{R}} \right\rbrace\nonumber\\
& -4\mu\left\lbrace\Tr \mathbf{\hat{Q}}_{A}\mathbf{\hat{Q}}_{A} \mathbf{R}\mathbf{R} -\Tr\mathbf{\hat{Q}}_{A}\mathbf{R}\mathbf{\hat{Q}}_{A} \mathbf{R} \right\rbrace
\label{Difference_MSEs_final}
\end{align}
Next, we will discuss equation \eqref{Difference_MSEs_final}. For
this purpose, we assume that the estimate of the projection matrix
of the signal subspace  $\mathbf{\hat{Q}}_{A}$ \cite{Vantrees}, the
true $\mathbf{R}$ \cite{Haykin} and the  data covariance matrices  $
\mathbf{\hat{R}} $ \cite{Haykin}  are Hermitian. For the next steps
we will make use of the following Theorem which is proved in
\cite{Chang}:

\medskip

\underline{Theorem 1}:
For two Hermitian matrices $\mathbf{A}$ and $\mathbf{B}$ of the same order,
\vspace{-0.5em}
\begin{align}
\label{Theorem_1}
& \Tr\left(\mathbf{A}\mathbf{B} \right)^{2^{k}}\leq \Tr\left(\mathbf{A}^{2^{k}}\mathbf{B}^{2^{k}} \right),
\end{align}
where k is in integer.\\

By replacing $\mathbf{A}$ with $\mathbf{\hat{Q}}_{A}$ and $\mathbf{B}$ with $\mathbf{\hat{R}}$ in \eqref{Theorem_1} and also considering $ k=1 $ , we have
\begin{align}
& \Tr\left(\mathbf{\hat{Q}}_{A}\mathbf{\hat{R}} \right)^{2}\leq \Tr\left(\mathbf{\hat{Q}}_{A}^{2}\mathbf{\hat{R}}^{2} \right)\nonumber\\
&\therefore\Tr\mathbf{\hat{Q}}_{A}\mathbf{\hat{R}} \mathbf{\hat{Q}}_{A}\mathbf{\hat{R}}\leq \Tr\mathbf{\hat{Q}}_{A}\mathbf{\hat{Q}}_{A}\mathbf{\hat{R}}\mathbf{\hat{R}}\nonumber\\
&\Rightarrow \Tr\mathbf{\hat{Q}}_{A}\mathbf{\hat{Q}}_{A}\mathbf{\hat{R}}\mathbf{\hat{R}}-\Tr\mathbf{\hat{Q}}_{A}\mathbf{\hat{R}} \mathbf{\hat{Q}}_{A}\mathbf{\hat{R}}\geq0
\label{Inequality_2}
\end{align}

\begin{figure}[!h]
    \centering 
    \includegraphics[width=8cm,height=6cm]{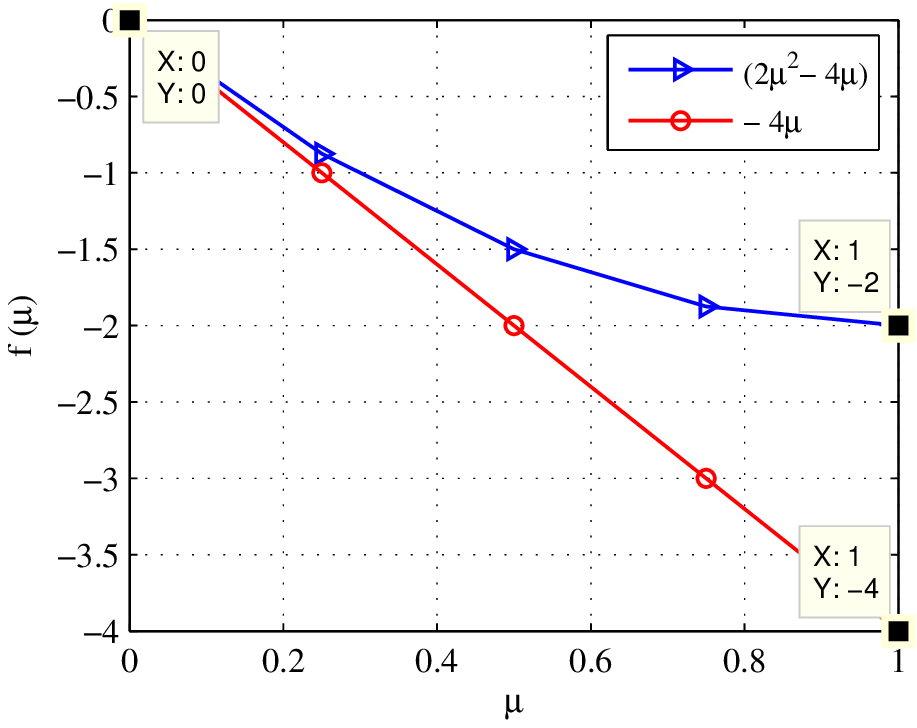} 
    \vspace{-1.0em}\caption{Behavior of $\left(\mathrm{2\mu}^{2}-\mathrm{4\mu}\right)$ and $-4\mu$ for $\mu$ $\in$ $[0\;1]$}
    \label{Behavior_reliability_factor4}
\end{figure}

Similarly, making $\mathbf{A}=\mathbf{\hat{Q}}_{A}$ and $\mathbf{B}=\mathbf{R}$ for $ k=1 $,  we obtain
\begin{align}
& \Tr\left(\mathbf{\hat{Q}}_{A}\mathbf{R} \right)^{2}\leq \Tr\left(\mathbf{\hat{Q}}_{A}^{2}\mathbf{R}^{2} \right)\nonumber\\
&\therefore\Tr\mathbf{\hat{Q}}_{A}\mathbf{R} \mathbf{\hat{Q}}_{A}\mathbf{R}\leq \Tr\mathbf{\hat{Q}}_{A}\mathbf{\hat{Q}}_{A}\mathbf{R}\mathbf{R}\nonumber\\&\Rightarrow \Tr\mathbf{\hat{Q}}_{A}\mathbf{\hat{Q}}_{A}\mathbf{R}\mathbf{R}-\Tr\mathbf{\hat{Q}}_{A}\mathbf{R} \mathbf{\hat{Q}}_{A}\mathbf{R}\geq0
\label{Inequality_3}
\end{align}
Next, we analyze the behavior of the expressions $-4\mu$ and
$\left(\mathrm{2\mu}^{2}-\mathrm{4\mu}\right)$ based on the
reliability factor $\mu $  $\in$ $[0\;1]$, as defined in
\eqref{modified_data_covariance}. In order to illustrate the case
being studied, we assume that both expressions are continuous
functions as depicted in Fig. \ref{Behavior_reliability_factor4}.
It can be seen in it that in the range $ [0\;1] $ both expressions assume values $\mathrm{f}(\mu)\leq0 $, i.e.:
\begin{align}
\mathrm{For}\:\mu\in[0\;1]: \left \{
\begin{aligned}
&\left(\mathrm{2\mu}^{2}-\mathrm{4\mu}\right)\leq0 \\
&-4\mu\leq0
\end{aligned} \right.
\label{Mu_conditions}
\end{align}
Now, we can consider the  traces which form the subtraction in  \eqref{Inequality_2} as different random variables $ \mathit{y}\left(\omega \right) $ and $ \mathit{x}\left(\omega \right) $, i.e.:
\begin{align}
\begin{aligned}
&\left.
\begin{aligned}
&\Tr\mathbf{\hat{Q}}_{A}\mathbf{\hat{Q}}_{A}\mathbf{\hat{R}}\mathbf{\hat{R}}=\mathit{y}\left(\omega \right)\\
&\Tr\mathbf{\hat{Q}}_{A}\mathbf{\hat{R}}\mathbf{\hat{Q}}_{A}\mathbf{\hat{R}}=\mathit{x}\left(\omega \right)
\end{aligned}
\right \},\:\forall\: \omega\:\in\Omega.
\end{aligned}
\label{Monotonicity_conditions_traces}
\end{align}
In addition, we can suppose that there is a random variable $
\mathit{z}\left(\omega \right) $ always greater than zero, i.e., $
\mathit{z}\left(\omega \right)\geq0 $, so that
\begin{align}
\mathit{z}\left(\omega \right) =\mathit{y}\left(\omega \right)-\mathit{x}\left(\omega \right)\geq0,\;\forall\: \omega\:\in\Omega
\label{Monotonicity_conditions_rv}
\end{align}
Taking the expectation of \eqref{Monotonicity_conditions_rv} and
applying its properties of linearity and monotonicity
\cite{Karr,JM}, we obtain
\begin{align}
&\mathbb{E}\left[  \mathit{z}\left(\omega \right)\right]
=\mathbb{E}\left[  \mathit{y}\left(\omega
\right)-\mathit{x}\left(\omega \right)\right] \geq0,
\label{Expect_Monotonicity_conditions_rv}
\end{align}
which, by making use of \eqref{Monotonicity_conditions_traces},
results in
\begin{align}
&\mathbb{E}\left[\mathit{z}\left(\omega \right)\right] =\mathbb{E}\left[  \mathit{y}\left(\omega \right)-\mathit{x}\left(\omega \right)\right]\nonumber\\
&=\mathbb{E}\left\lbrace\Tr\mathbf{\hat{Q}}_{A}\mathbf{\hat{Q}}_{A}\mathbf{\hat{R}}\mathbf{\hat{R}}  -\Tr\mathbf{\hat{Q}}_{A}\mathbf{\hat{R}}\mathbf{\hat{Q}}_{A}\mathbf{\hat{R}} \right\rbrace \geq0
\label{Expect_Monotonicity_conditions_trace}
\end{align}
Next, we can combine the inequalities \eqref{Mu_conditions} with
\eqref{Expect_Monotonicity_conditions_trace} to compute the  second
member of \eqref{Difference_MSEs_final}, for $  \mu\in[0\;1]$.

For its first summand, we combine \eqref{Mu_conditions} and
\eqref{Expect_Monotonicity_conditions_trace}, as follows:
\begin{align}
\left\lbrace
\begin{aligned}
&\mathbb{E}\left\lbrace\Tr\mathbf{\hat{Q}}_{A}\mathbf{\hat{Q}}_{A}\mathbf{\hat{R}}\mathbf{\hat{R}}  -\Tr\mathbf{\hat{Q}}_{A}\mathbf{\hat{R}}\mathbf{\hat{Q}}_{A}\mathbf{\hat{R}} \right\rbrace \geq0\\
&\left(\mathrm{2\mu}^{2}-\mathrm{4\mu}\right)\leq0,\;\mu\in[0\;1],
\end{aligned}
\right.
\label{Proof_MSE_final_firstsummand_1}
\end{align}
to obtain in a straightforward way
\begin{align}
&\left(\mathrm{2\mu}^{2}-\mathrm{4\mu}\right)\mathbb{E}\left\lbrace\Tr\mathbf{\hat{Q}}_{A}\mathbf{\hat{Q}}_{A}\mathbf{\hat{R}}\mathbf{\hat{R}}  -\Tr\mathbf{\hat{Q}}_{A}\mathbf{\hat{R}}\mathbf{\hat{Q}}_{A}\mathbf{\hat{R}} \right\rbrace \leq0
\label{Proof_MSE_final_first summand_2}
\end{align}
Similarly, we can compute its second member, by combining
\eqref{Mu_conditions} and \eqref{Inequality_3}, as described by
\begin{align}
\left\lbrace
\begin{aligned}
&\Tr\mathbf{\hat{Q}}_{A}\mathbf{\hat{Q}}_{A}\mathbf{R}\mathbf{R}  -\Tr\mathbf{\hat{Q}}_{A}\mathbf{\hat{R}}\mathbf{\hat{Q}}_{A}\mathbf{\hat{R}}  \geq0\\
&-\mathrm{4\mu}\leq0,\;\mu\in[0\;1],
\end{aligned}
\right.
\label{Proof_MSE_final_second_summand_1}
\end{align}
to obtain also straightforwardly the expression given by
\begin{align}
&-\mathrm{4\mu}\left\lbrace\Tr\mathbf{\hat{Q}}_{A}\mathbf{\hat{Q}}_{A}\mathbf{R}\mathbf{R}  -\Tr\mathbf{\hat{Q}}_{A}\mathbf{R}\mathbf{\hat{Q}}_{A}\mathbf{R} \right\rbrace \leq0
\label{Proof_MSE_final_second_summand_2}
\end{align}
By combining the inequalities \eqref{Proof_MSE_final_first
    summand_2} and \eqref{Proof_MSE_final_second_summand_2} with
\eqref{Difference_MSEs_final}, we have
\begin{align}
&\mathrm{MSE}\left( \mathbf{\hat{R}}^{(n+1)}\right)\big\rvert_{n=1}-\mathrm{MSE}\left( \mathbf{\hat{R}}\right)\nonumber\\
&=\underbrace{\left( \mathrm{2\mu}^{2}-\mathrm{4\mu}\right)  \mathbb{E}\left\lbrace\Tr\mathbf{\hat{Q}}_{A}\mathbf{\hat{Q}}_{A}\mathbf{\hat{R}}\mathbf{\hat{R}} -\Tr\mathbf{\hat{Q}}_{A}\mathbf{\hat{R}}\mathbf{\hat{Q}}_{A}\mathbf{\hat{R}} \right\rbrace}_{\leq\;0}\nonumber\\
&\underbrace{-4\mu\left\lbrace\Tr \mathbf{\hat{Q}}_{A}\mathbf{\hat{Q}}_{A} \mathbf{R}\mathbf{R} -\Tr\mathbf{\hat{Q}}_{A}\mathbf{R}\mathbf{\hat{Q}}_{A} \mathbf{R} \right\rbrace}_{\leq\;0}\\
& \therefore\;\mathrm{MSE}\left( \mathbf{\hat{R}}^{(n+1)}\right)\big\rvert_{n=1}-\mathrm{MSE}\left( \mathbf{\hat{R}}\right)\leq\:0
\label{Proof_final}
\end{align}
which is the desired result.

\end{document}